\documentclass[aps,pre,twocolumn,nopacs,final,letterpaper,superscriptaddress,longbibliography]{revtex4-1}

\usepackage[utf8]{inputenc}
\usepackage{calc}
\usepackage{graphicx}
\usepackage{amsmath,amssymb,amsthm}
\usepackage{mathtools}
\usepackage{txfonts}
\usepackage{bm}
\usepackage{color}
\usepackage{hyperref}
\usepackage{multirow}
\graphicspath{{./figs/}}

\begin{document}
\author{Guillaume St-Onge}
\affiliation{D\'epartement de physique, de g\'enie physique et d'optique,
Universit\'e Laval, Qu\'ebec (Qu\'ebec), Canada G1V 0A6}
\affiliation{Centre interdisciplinaire en mod\'elisation math\'ematique, Universit\'e Laval, Qu\'ebec (Qu\'ebec), Canada G1V 0A6}
\author{Vincent Thibeault}
\affiliation{D\'epartement de physique, de g\'enie physique et d'optique,
Universit\'e Laval, Qu\'ebec (Qu\'ebec), Canada G1V 0A6}
\affiliation{Centre interdisciplinaire en mod\'elisation math\'ematique, Universit\'e Laval, Qu\'ebec (Qu\'ebec), Canada G1V 0A6}
\author{Antoine Allard}
\affiliation{D\'epartement de physique, de g\'enie physique et d'optique,
Universit\'e Laval, Qu\'ebec (Qu\'ebec), Canada G1V 0A6}
\affiliation{Centre interdisciplinaire en mod\'elisation math\'ematique, Universit\'e Laval, Qu\'ebec (Qu\'ebec), Canada G1V 0A6}
\author{Louis J. Dub\'{e}}
\affiliation{D\'epartement de physique, de g\'enie physique et d'optique,
Universit\'e Laval, Qu\'ebec (Qu\'ebec), Canada G1V 0A6}
\affiliation{Centre interdisciplinaire en mod\'elisation math\'ematique, Universit\'e Laval, Qu\'ebec (Qu\'ebec), Canada G1V 0A6}
\author{Laurent H\'{e}bert-Dufresne}
\affiliation{D\'epartement de physique, de g\'enie physique et d'optique,
Universit\'e Laval, Qu\'ebec (Qu\'ebec), Canada G1V 0A6}
\affiliation{Vermont Complex Systems Center, University of Vermont, Burlington, VT 05405 }
\affiliation{Department of Computer Science, University of Vermont, Burlington, VT 05405 }

\title{Master equation analysis of mesoscopic localization in contagion dynamics \\ on higher-order networks}

\begin{abstract}
Simple models of infectious diseases tend to assume random mixing of individuals, but real interactions are not random pairwise encounters: they occur within various types of gatherings such as workplaces, households, schools, and concerts, best described by a higher-order network structure. We model contagions on higher-order networks using group-based approximate master equations, in which we track all states and interactions within a group of nodes and assume a mean-field coupling between them. Using the Susceptible-Infected-Susceptible dynamics, our approach reveals the existence of a mesoscopic localization regime, where a disease can concentrate and self-sustain only around large groups in the network overall organization. In this regime, the phase transition is smeared, characterized by an inhomogeneous activation of the groups. At the mesoscopic level, we observe that the distribution of infected nodes within groups of a same size can be very dispersed, even bimodal. When considering heterogeneous networks, both at the level of nodes and groups, we characterize analytically the region associated with mesoscopic localization in the structural parameter space. We put in perspective this phenomenon with eigenvector localization and discuss how a focus on higher-order structures is needed to discern the more subtle localization at the mesoscopic level. Finally, we discuss how mesoscopic localization affects the response to structural interventions and how this framework could provide important insights for a broad range of dynamics.
\end{abstract}

\maketitle

\section{Introduction}

Classic epidemiological models have been successful at providing meaningful insights on the spreading of infection diseases \cite{anderson1992infectious,diekmann1995legacy}.
Their simplicity is their strength : from as little information as the basic reproduction number $R_0$, one can tell whether or not a disease should invade or not a population.
However, we cannot hope to represent the complexity of human behavior and of our modern social structure with mathematical models relying solely on an average individual.
This is even more true when considering more complex types of spreading processes, such as social contagions \cite{Centola2007,Monsted2017,Lehmann2018} or the coevolution of diseases \cite{Hebert-Dufresne2015}.

The study of spreading processes on networks allows to look beyond the mass action principle, to account for more realistic contact patterns while keeping our models simple enough to provide meaningful insights \cite{Pastor-Satorras2015,Kiss2017}.
One success of network science has been to unveil the impact of the heterogeneity of contacts, and how this affects critical properties of these systems.
Heterogeneous mean-field theories \cite{Pastor-Satorras2001,Boguna2002}, heterogeneous pair approximations \cite{Eames2002,Mata2014}, and approximate master equations \cite{marceau2010adaptive,Gleeson2011,Lindquist2011} represent only a few of the many techniques that have been developed to describe the behavior of dynamical processes on networks with heterogeneous number of contacts.

Social networks, however, are more than just random contacts between heterogeneous individuals: interactions occur in a coordinated manner because of a higher-level organization.
At the mesoscopic level, we see groups of individuals that are more or less densely connected to one another \cite{girvan2002community,Newman2003}.
We can thus shift from asking \textit{if} a contagion can invade a population, to \textit{where} it should thrive within that population. 
This question is best embodied by the phenomenon of \textit{epidemic localization}: near the epidemic threshold, the disease exists only in some parts of the whole network.

The localization of epidemics has been studied mostly through the lens of extensive numerical simulations or quenched mean-field theory \cite{Goltsev2012,Castellano2012,Pastor-Satorras2018,Liu2019}.
A general observation is that for most complex networks, an epidemic should either be localized around the innermost network core or the hubs \cite{Castellano2012}.
The localization subgraph depends on the structure, but also on the details of the dynamics \cite{Ferreira2016,Cota2018robustness}.
Moreover, localization dramatically affects the fundamental critical properties of an epidemic : it is notably possible to observe a Griffiths phase, where the system slowly relaxes to an inactive state \cite{odor2015griffiths,Cota2018,Vojta2006,Odor2014,Cota2016}. Another notable effect is the smearing of the phase transition, where the order parameter develops inhomogeneously beyond the critical point \cite{Vojta2006,Odor2014,Cota2016,St-Onge2018,hebert2019smeared}.

Despite the important body of work on epidemic localization, theoretical results are still limited to a handful of models and most works have used a node-centric perspective.
To broaden our understanding of localization of dynamical processes and embrace the higher-level organization of complex networks, we argue that higher-order representations of networks should be used \cite{Battiston2020}.
Furthermore, we claim that approximate master equations represent powerful and flexible approaches for this purpose.

\begin{figure}[tb]
\centering
\includegraphics[width=\linewidth]{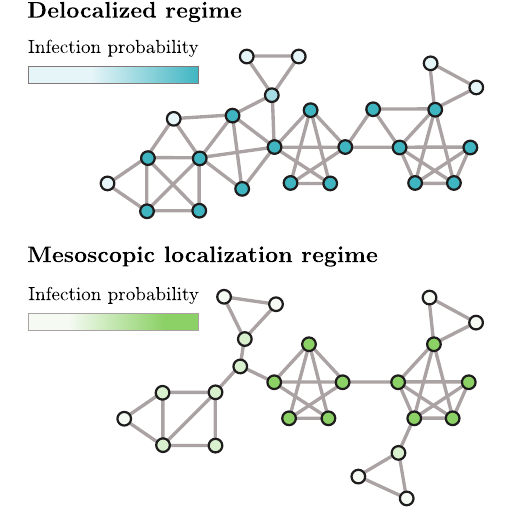}
\caption{Simple illustration of the mesoscopic localization phenomenon. In both regimes, the contagion is concentrated around the innermost core of the network, but the composition of the core is different. In what we called the delocalized regime, substructures of all sizes (e.g. triangles, $4-$cliques, etc.) contribute to the contagion, while in the mesoscopic localization regime, there is a strong bias toward the largest and densest substructures.}
\label{fig:toy_viz}
\end{figure}

In this paper, we reveal a phenomenon we call \textit{mesoscopic localization} for spreading processes on certain heterogeneous networks with a higher-level organization (see Fig.~\ref{fig:toy_viz}).
It is characterized by the localization of the contagion in large but finite-size mesoscopic substructures near the epidemic threshold \footnote{Substructures at the mesoscopic level are meant to describe a large number of nodes, organized in more or less compact groups, but are by nature non-extensive. They could stand for schools, households, sports teams, etc. in a contact network.}, with a phase transition that is smeared at the global level.
To illustrate this phenomenon, we use a group-based framework together with an approximate master equation analysis of the SIS dynamics.
We present a complete analytical description of the mesocopic localization regime, while we describe its impact on interventions in Ref.~\cite{st2020school} to show how accounting for this localization regime is critical to our ability to control contagions on networks.

This paper is structured as follows.
First, we introduce the group-based framework and the approximate master equations in Sec.~\ref{sec:theoretical_framework}.
We obtain an implicit expression and explicit bounds for the epidemic threshold  in Sec.~\ref{subsec:epidemic_threshold}.
With a development of the stationary state near the critical point, we show in Sec.~\ref{subsec:heterogeneous_structure} that mesoscopic localization emerges from a sufficiently weak coupling between the groups.
Second, we fully characterize mesoscopic localization in Sec.~\ref{sec:mesoscopic_localization}. We derive asymptotic results for the scaling of the epidemic threshold in Sec.~\ref{subsec:localization_regimes}, leading to explicit expressions for the localization regimes. We then consider the effects of finite-size cut-offs in Sec.~\ref{subsec:finite_cutoffs}.
We complete our analysis using the \textit{inverse participation ratio}, further connecting our work with the literature on eigenvector localization.
Our comparison reveals the importance of a change of perspective---a focus on higher-level group organization rather than individual nodes---in order to detect localization phases at the mesoscopic level.
Finally, in Sec.~\ref{sec:discussion}, we discuss possible extensions of our work and some direct implications for the control of epidemics \cite{st2020school}.

\section{Group-based SIS model}
\label{sec:theoretical_framework}

\begin{figure}[tb]
\centering
\includegraphics[width=\linewidth]{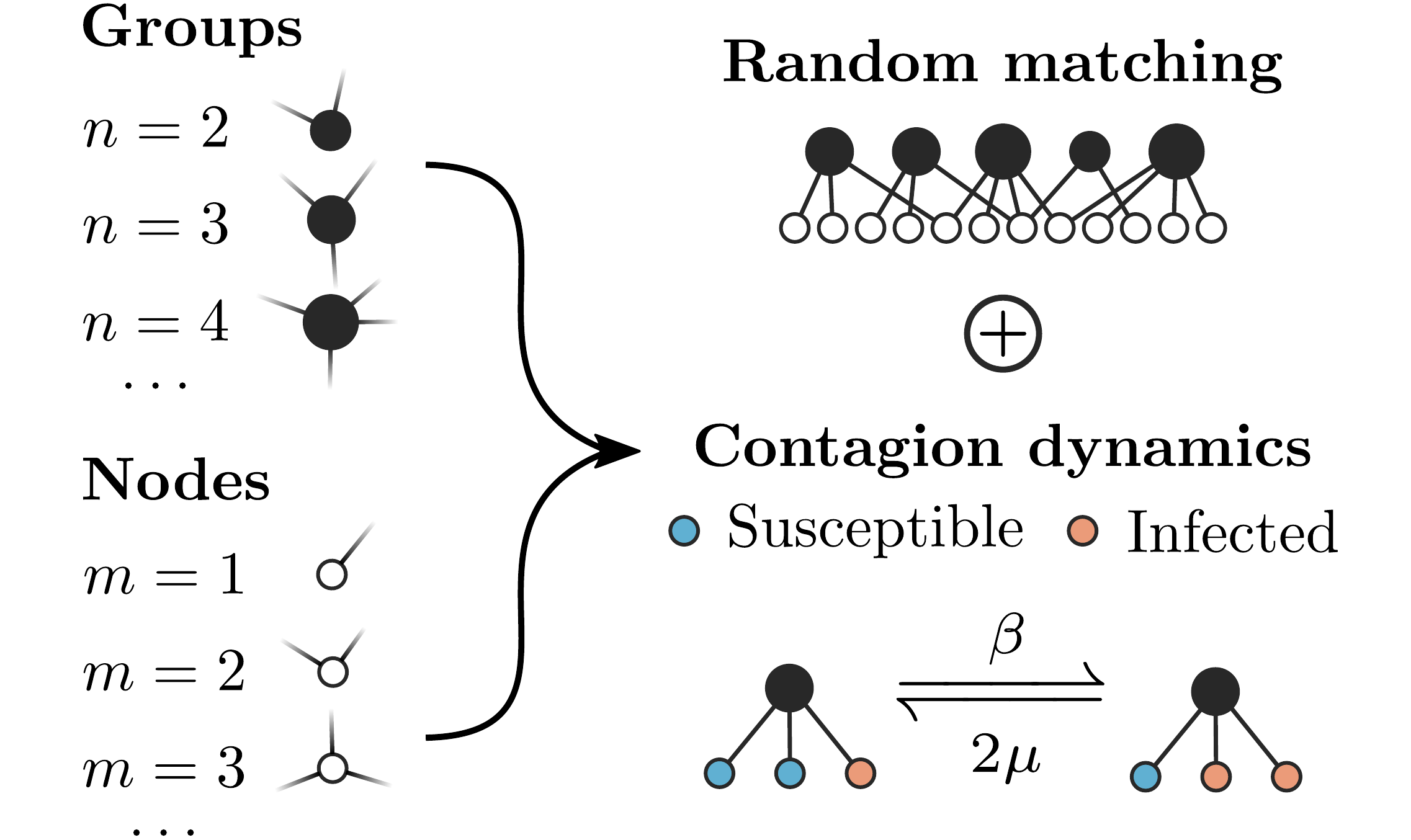}
\caption{Framework for contagions on higher-order networks. Nodes are assigned to $m$ groups and groups are of various sizes $n$, distributed according to $g_m$ and $p_n$. Groups are equivalent to cliques in the main text. We consider a SIS dynamics where infected nodes transmit the disease at rate $\beta$, and recover at rate $\mu$.}
\label{fig:framework}
\end{figure}

There exists multiple representations for higher-order structures \cite{Battiston2020}, ranging from simplicial complexes to hypergraphs, or more simply a bipartite graph, where nodes are attached to \textit{groups} that encode the interaction.
In this paper, we use the latter [see Fig.~\ref{fig:framework}].
Groups could be used to represent any kind of mesoscopic substructures, dense or sparse, with possibly weighted and directed edges.
They could also be used to model higher-order interactions that cannot be decomposed into simpler pairwise interactions.
To simplify the mathematical description, in the main text we consider that all groups of $n$ nodes represent \textit{cliques}, i.e., fully connected and undirected subgraphs.
In Appendix \ref{app:scaling_transmission}, we generalize the approach to consider weighted cliques.

We assume that each node in the network belongs to a certain number of groups, $m$, the \textit{membership} of the node, which is drawn from a membership distribution $g_m$.
The size $n$ of a group is drawn from a group size distribution $p_n$.
We consider infinite-size heterogeneous random networks where nodes are assigned to groups uniformly at random \cite{Newman2003}.
In other words, $m$ and $n$ are uncorrelated.
Throughout the paper, we denote expected values taken over $p_n$ and $g_m$ as $\left  \langle \, \cdots \right \rangle$, where the interior of the bracket makes it clear over which distribution the average is performed.

Let us introduce a few structural properties associated with this ensemble.
The average membership of a node is $\langle m \rangle$ and the average group size is $\langle n \rangle$.
If we pick a node at random and follow a group to which it belongs, the distribution for the size of that group is proportional to $n p_n$. Consequently, the average \textit{excess} group size, i.e., the average number of neighbors this node has in that group, is $\left \langle n(n-1) \right \rangle/\langle n \rangle$.
Since $m$ and $n$ are uncorrelated, the average degree of a node (in the one-mode projection of the bipartite graph) is therefore
\begin{align*}
    \frac{\langle m \rangle \left \langle n(n-1) \right \rangle}{\langle n \rangle} \;.
\end{align*}

On these networks, we consider the Susceptible-Infected-Susceptible (SIS) dynamics in which each node is either infected or susceptible.
Infected nodes transmit the disease to their neighbors at rate $\beta$ and recover to the susceptible state at a rate $\mu$ [see Fig.~\ref{fig:framework}].

We describe the dynamics using the heterogeneous clique approximation of Ref.~\cite{hebert2010propagation}.
We track $s_m(t)$, the probability for a node of membership $m$ to be susceptible at time $t$, and $c_{n,i}(t)$, the probability to observe $i$ infected nodes within a group of size $n$ at time $t$.

We define the following system of approximate master equations
\begin{subequations}
\label{eq:ame_ode}
\begin{align}
    \frac{\mathrm{d}s_m}{\mathrm{d}t} =& \;\mu (1 - s_m) - m r s_m  \;, \\
    \frac{\mathrm{d}c_{n,i}}{\mathrm{d}t} =& \;\mu (i+1) c_{n,i+1} - \mu i c_{n,i} +  (n-i+1) \lbrace \beta(i-1) + \rho\rbrace c_{n,i-1} \notag  \\
    &-  (n-i) \lbrace \beta i + \rho\rbrace c_{n,i} \;, \label{eq:ame_ode_cni}
\end{align}
\end{subequations}
which contains a total of $\mathcal{O}(m_\mathrm{max} + n_\mathrm{max}^2 )$ equations, where $m_\mathrm{max}$ and $n_\mathrm{max}$ are the maximal membership and maximal group size respectively.
From now on, we set $\mu \equiv 1$ without loss of generality.

The mean fields $r(t)$ and $\rho(t)$ are defined as
\begin{subequations}
  \label{eq:mean_field}
\begin{align}
    r(t) &= \frac{  \sum_{n,i} \beta i \;(n-i)c_{n,i}(t) p_n}{\sum_{n,i} (n-i) c_{n,i}(t) p_n} \;, \label{eq:mean-field1} \\
    \rho(t) &= r(t) \left [ \frac{\sum_m (m-1)\; ms_m(t) g_m}{ \sum_m m s_m(t) g_m} \right ] \;. \label{eq:mean-field2}
\end{align}
\end{subequations}
If we take a susceptible node and select a random group to which it belongs, $r(t)$ is the mean infection rate associated to that group. Indeed, the joint distribution for the size $n$ of the group and the number of infected nodes $i$ within that group is proportional to \mbox{$(n-i) c_{n,i}(t) p_n$}. Then $r(t)$ is just an average of the infection rate received, $\beta i$, over this joint distribution.

Now if we pick a susceptible node in a group, $\rho(t)$ is the mean infection rate received from all external groups (i.e., excluding the one we picked the node from).
Assuming that infection coming from different groups are independent processes, we multiply $r(t)$ with the mean \textit{excess} membership of a susceptible node to get $\rho(t)$.
The membership distribution of a susceptible node picked in a group is proportional to $m s_m(t) g_m$, thus we simply average $m-1$, the excess membership, over this distribution.

The global prevalence (average fraction of infected nodes) is
\begin{align*}
    I(t) = \sum_m g_m [1-s_m(t)] \;,
\end{align*}
and the prevalence within groups of size $n$ is
\begin{align*}
    I_n(t)  = \sum_{i} \frac{i}{n} \; c_{n,i}(t) \;.
\end{align*}
Note that unless specified otherwise, sums over $m$ ($n$) are over every value such that $g_m > 0$ ($p_n > 0$), and sums over $i$ cover the range $\lbrace 0, \dots, n \rbrace$.

In Eq.~\eqref{eq:ame_ode}, the evolution of each $s_m$ is treated in a mean-field fashion~\footnote{Note that we are still preserving the dynamic correlations between pairs of nodes by tracking each $c_{n,i}$.}, while the evolution of each $c_{n,i}$ is described using a master equation. The infection rate due to infected nodes \textit{within} a group is treated exactly, while the contribution of infected nodes in \textit{external} groups is approximated (i.e. the terms involving $\rho$). We therefore refer to our approach as \textit{approximate master equations}.

The system eventually settles to a stationary state in the limit $t \to \infty$, and henceforth we assume that the quantities $s_m, c_{n,i}, r$ and $\rho$ have reached a fixed point. These variables characterizing the stationary state are obtained by solving the following self-consistent expressions
\begin{subequations}\label{eq:stationary_state}
\begin{align}
    s_m =& \frac{1}{1 + m r} \;, \label{eq:stationary_state_sm}\\
    (i+1) c_{n,i+1} =& \lbrace i + (n-i) \left [ \beta i + \rho \right ] \rbrace c_{n,i} \;, \notag \\
                     &- (n-i+1)\left [ \beta (i-1) + \rho \right ] c_{n,i-1} \;, \label{eq:stationary_state_fni}
\end{align}
\end{subequations}
obtained from Eq.~\eqref{eq:ame_ode}, and where $r$ and $\rho$ are still given by Eq.~\eqref{eq:mean_field}. It will be useful to rewrite Eq.~\eqref{eq:stationary_state_fni} more explicitly as
\begin{align}
    \label{eq:cni_explicit}
    c_{n,i} = c_{n,0} \frac{n!}{(n-i)!i!} \prod_{j=0}^{i-1} [ \beta j + \rho] \quad \forall i \in \lbrace 1, \dots, n \rbrace \;,
\end{align}
with $c_{n,0} = 1-\sum_{i=1}^n c_{n,i}$.

\subsection{Epidemic threshold}
\label{subsec:epidemic_threshold}

For the SIS dynamics, there exists a critical value $\beta_\mathrm{c}$ for the transmission rate, called the epidemic threshold. For \mbox{$\beta < \beta_\mathrm{c}$}, the absorbing-state---where all nodes are susceptible---is attractive for all initial conditions. For \mbox{$\beta > \beta_\mathrm{c}$}, the absorbing-state becomes unstable and there exists a non-trivial stationary state.

To obtain an expression for $\beta_\mathrm{c}$, let us redefine the stationary state observables as functions of $\rho$, i.e., $r(\rho)$, $s_m(\rho)$ and $c_{n,i}(\rho)$.
We then define the right-hand side of Eq.~\eqref{eq:mean-field2} as $F(\rho)$. Since $F(\rho)$ is bounded from above \footnote{In Eqs.~(\ref{eq:mean_field}a-b), $c_{n,i}$ and $s_m$ remain bounded for all $\rho$.}, a positive solution \mbox{$\rho = F(\rho)$} exists if
\begin{align*}
    \left. \frac{\mathrm{d} F}{\mathrm{d} \rho}  \right|_{\rho \to 0} > 1 \;.
\end{align*}
At the epidemic threshold $\beta_\mathrm{c}$, this derivative is exactly $1$, resulting in $\rho \to 0$, $r(\rho) \to 0$, $s_m(\rho) \to 1$ and $c_{n,i}(\rho) \to \delta_{i,0}$, where $\delta_{i,j}$ is the Kronecker delta.

It will prove useful to expand $c_{n,i}$ near the epidemic threshold as $c_{n,i}(\rho) = \delta_{i,0} + h_{n,i} \rho + \mathcal{O}(\rho^2)$.
From Eq.~\eqref{eq:cni_explicit}, we obtain
\begin{align*}
    h_{n,i} \equiv \left. \frac{\mathrm{d} c_{n,i}}{\mathrm{d}\rho} \right|_{\rho \to 0}  = \frac{n! \beta^{i-1} (i-1)!}{(n-i)! i!} \quad \forall i \in \lbrace 1, \dots, n \rbrace \;,
\end{align*}
and by definition $h_{n,0} \equiv - \sum_{i = 1}^n h_{n,i}$.

For all $n$, we encode each sequence $\left ( h_{n,i} \right)_{i = 0}^n$ in the generating function
\begin{align}
    H_n(x;\beta) &= \sum_{i} h_{n,i} x^i \;, \notag \\
           &= h_{n,0} + \frac{1}{\beta} \sum_{i = 1}^n \frac{n!}{(n-i)!i!} (\beta x)^i (i-1)! \;, \notag \\
           &= h_{n,0} + \frac{1}{\beta} \int_0^\infty \sum_{i = 1}^n \frac{n!}{(n-i)!i!} (\beta u x)^i u^{-1} e^{-u} \mathrm{d} u  \;,\notag \\
           &= h_{n,0} + \frac{1}{\beta} \int_0^\infty  [(1 + \beta u x)^n - 1] u^{-1} e^{-u} \mathrm{d} u  \;. \label{eq:gen_func_Hn}
\end{align}
Interestingly, the auxiliary generating function
\begin{align*}
    Q_n(x;\beta) = \frac{H_n(x;\beta) - h_{n,0}}{H_n(1;\beta) - h_{n,0}} = \frac{\int_0^\infty  [(1 + \beta u x)^n - 1] u^{-1} e^{-u} \mathrm{d} u}{\int_0^\infty  [(1 + \beta u)^n - 1] u^{-1} e^{-u} \mathrm{d} u} \;,
\end{align*}
can be interpreted as the probability generating function for the \textit{quasi-stationary distribution} (only for \mbox{$i>0$}) of the number of infected nodes in a group of size $n$, under the influence of a weak (vanishing) external field.

These generating functions allow to write
\begin{align}
\label{eq:pre_epidemic_threshold}
\left. \frac{\mathrm{d} F}{\mathrm{d} \rho}  \right|_{\rho \to 0} = \beta \frac{\langle m(m-1) \rangle}{\langle m \rangle  \langle n \rangle } \left \langle (n-1) H_n'(1;\beta) -  H_n''(1;\beta) \right \rangle \;,
\end{align}
where the derivatives are with respect to $x$ and we have used standard properties of generating functions in combination with Eqs.~\eqref{eq:mean-field1} and \eqref{eq:mean-field2}.
We simplify the above equation by noting that
\begin{align*}
    (n-1)H_n'(1;\beta) - H_n''(1;\beta) &= n(n-1) \int_0^\infty (1+\beta u)^{n-2} e^{-u} \mathrm{d} u \;.
\end{align*}
The epidemic threshold $\beta_\mathrm{c}$ is thus obtained by solving the following implicit equation for $\beta$
\begin{align}\label{eq:epidemic_threshold}
    \beta \frac{\langle m(m-1) \rangle}{\langle m \rangle  \langle n \rangle } \left \langle n(n-1) A_n(\beta) \right \rangle  = 1\;,
\end{align}
where
\begin{align}
    \label{eq:An}
    A_n(\beta) \equiv \int_0^\infty (1+\beta u)^{n-2} e^{-u} \mathrm{d} u \;.
\end{align}
Appendix \ref{app:epidemic_dev} provides a detailed development leading to Eqs.~\eqref{eq:pre_epidemic_threshold}-\eqref{eq:An}.
$A_n$ can also be rewritten in terms of the upper incomplete gamma function, but the present integral representation will be more useful later on.

Although it is not possible to write $\beta_\mathrm{c}$ in closed form, we provide bounds for $A_n(\beta)$,
\begin{align*}
    1 \leq  A_n(\beta) \leq  \frac{1}{1-\beta(n_\mathrm{max}-2)} \;.
\end{align*}
Details of this result are presented in Appendix \ref{app:bounds_threshold}. These inequalities lead to lower and upper bounds on the epidemic threshold
\begin{subequations}
\label{eq:bounds_threshold}
\begin{align}
    \beta_\mathrm{c} &\geq \frac{1}{\Omega(g_m,p_n) + (n_\mathrm{max} - 2)} \;, \label{eq:lower_bound_threshold}\\
    \beta_\mathrm{c} &\leq \frac{1}{\Omega(g_m,p_n)} \label{eq:upper_bound_threshold}\;,
\end{align}
\end{subequations}
where we defined the \textit{coupling} between groups as
\begin{align}
    \label{eq:coupling}
    \Omega(g_m,p_n) \equiv \left (\frac{\langle m(m-1) \rangle}{\langle m \rangle} \right) \left ( \frac{ \langle n(n-1) \rangle}{ \langle n \rangle} \right) \;,
\end{align}
the product of the average excess group size and the average excess membership.
If we take a random node within a group, $\Omega(g_m,p_n)$ corresponds to its average number of \textit{external neighbors}.
It is therefore a good measure of the interaction of groups with one another.

\subsection{Behavior for heterogeneous membership and group size}
\label{subsec:heterogeneous_structure}

\begin{figure*}[tb]
\centering
\includegraphics[width=\linewidth]{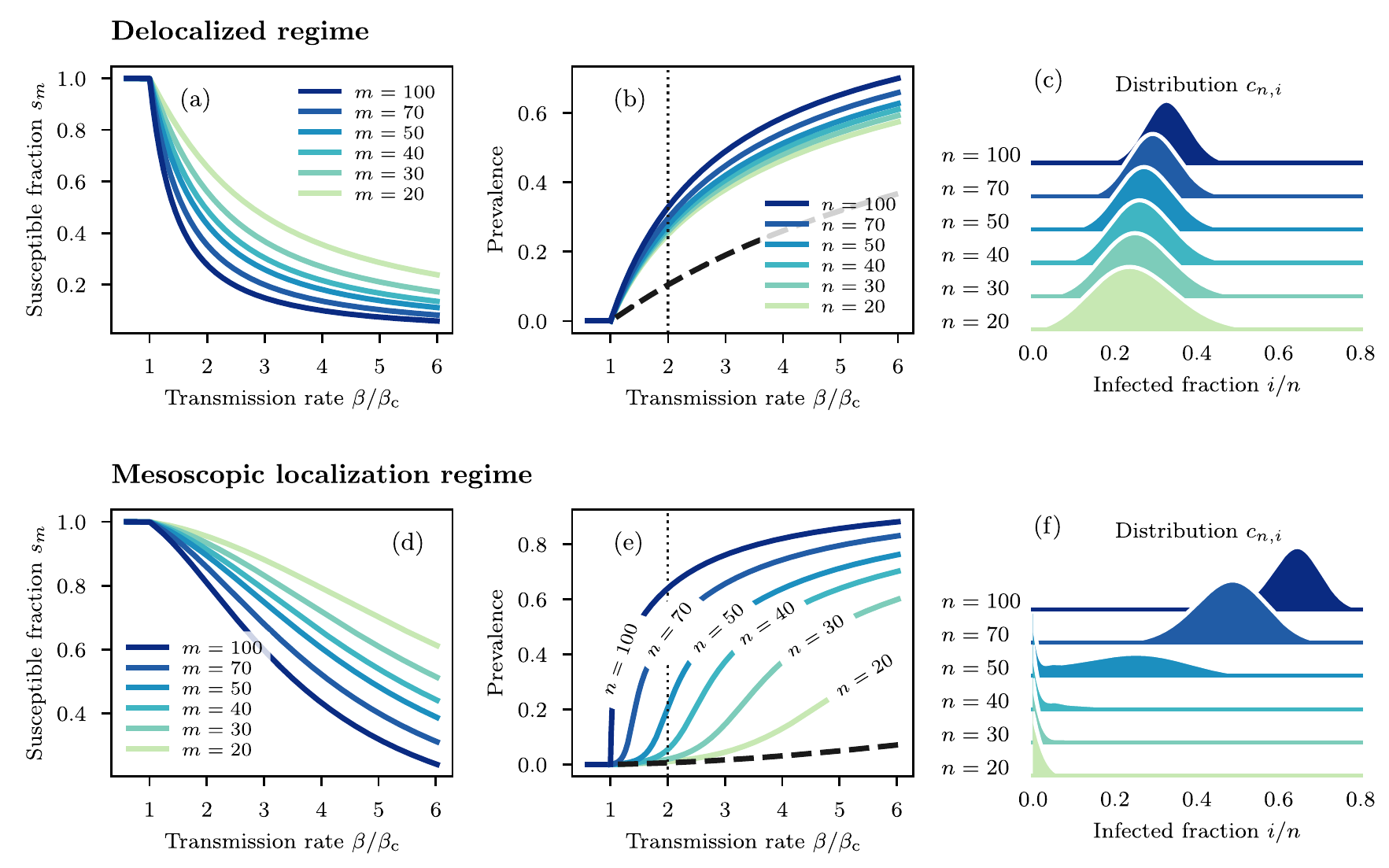}
\caption{Comparison of the stationary state near the epidemic threshold in the delocalized and mesoscopic localization regimes. Stationary state solutions were obtained from Eqs.~\eqref{eq:stationary_state_sm} and \eqref{eq:cni_explicit} for heterogeneous membership and group size distributions of the form $g_m \propto m^{-\gamma_m}$ and $p_n \propto n^{-\gamma_n}$. We used $m,n \in \lbrace 2,\dots,100 \rbrace$. The epidemic threshold $\beta_\mathrm{c}$ is the solution to Eq.~\eqref{eq:epidemic_threshold}. (a) and (d) Stationary fraction of susceptible nodes with a given membership $m$ as a function of the transmission rate. (b) and (e) Group prevalence $I_n$ (solid lines) and global prevalence $I$ (dashed line) as a function of the transmission rate. (c) and (f) Distributions for the number of infected nodes $i$ in a group of size $n$, obtained for $\beta = 2 \beta_\mathrm{c}$ corresponding to the vertical dotted lines in the panel on their left. Spline interpolations are used for visual purpose. Upper row (a)-(c) $\gamma_m = \gamma_n = 2.2$. Lower row (d)-(f) $\gamma_m = 4$ and $\gamma_n = 3.5$. }
\label{fig:local}
\end{figure*}

Let us consider power-law distributions $p_n \propto n^{-\gamma_n}$ and \mbox{$g_m \propto m^{-\gamma_m}$} with large cut-offs $n_\mathrm{max} \gg 1$ and $m_\mathrm{max} \gg 1$.
We set $\gamma_n, \gamma_m > 2$ so that $\langle n \rangle$ and $\langle m \rangle$ remain bounded.

For reasons that will become clear in Sec.~\ref{sec:mesoscopic_localization}, we distinguish a \textit{strong} group coupling ($\Omega(g_m,p_n) \gg n_\mathrm{max}$) from a \textit{weak} group coupling ($\Omega(g_m,p_n) \ll n_\mathrm{max}$).
Figure~\ref{fig:local} illustrates the stationary state properties of the dynamics for two different pairs of exponents $(\gamma_m,\gamma_n)$, Fig.~\ref{fig:local}(a-c) corresponding to a strong group coupling and Fig.~\ref{fig:local}(d-f) to a weak group coupling.

Comparing Fig.~\ref{fig:local}(a) and Fig.~\ref{fig:local}(d), we note that all $s_m$ decrease faster in the former case as the ratio $\beta/\beta_\mathrm{c}$ increases. From Eq.~\eqref{eq:stationary_state_sm}, this is explained by a faster increase of the mean field $r$, resulting directly from a stronger coupling between groups.

The difference between Fig.~\ref{fig:local}(b)~and~\ref{fig:local}(e) is more striking.
While the group prevalence $I_n$ does not vary much with $n$ in Fig.~\ref{fig:local}(b)---the coupling $\Omega(g_m,p_n)$ is strong---we observe a sequential activation of the groups for the weakly coupled system in Fig.~\ref{fig:local}(e).
Figures~\ref{fig:local}(c)~and~\ref{fig:local}(f) provide an even clearer illustration for a fixed $\beta$.
When the coupling is strong, all distributions $c_{n,i}$ are concentrated around roughly the same fraction of infected nodes within the groups.  Weak coupling yields a more diverse scenario where smaller groups have very few infected nodes while the prevalence in larger groups can be very high.  We qualify the latter as \textit{active} groups.
For groups of moderate size (e.g., $n=50$), $c_{n,i}$ is bimodal and highly dispersed, akin to a system near a critical point.

This is a telling illustration of why stochastic dynamics on networks with a high level of group organization are best described by approximate master equations: groups of nodes can have heterogeneous state distributions, and a cruder approximation (e.g., models averaging $i/n$ for all groups of a given size or other mean-field approximations) is likely to miss many rich features of the dynamics.
These features may be interesting by themselves, and important for the overall evolution of the process.
While mean-field approaches are sometime \textit{qualitatively} correct, they are most often \textit{quantitatively} off the mark \cite{Gleeson2011,Gleeson2013}.
Approximate master equations yield both qualitatively and quantitatively correct results (see Appendix \ref{app:MC_validation}), ensuring that the observed phenomena are true properties of the original stochastic process.

The scenario presented by Figs.~\ref{fig:local}(e)~and~\ref{fig:local}(f) is typical of a \textit{smeared phase transition}.
Instead of clean critical point driven by a collective ordering, subparts of the system self-activate independently from the rest, as shown by the local order parameters $I_n$.
This behavior has an intuitive explanation.
Since $p_n \propto n^{-\gamma_n}$, a small proportion of the groups are very large, albeit of finite size.
Near $\beta_\mathrm{c}$, the largest groups are able to self-sustain an endemic state by themselves, but since the coupling is weak, the contagion does not spread through the rest of the network.
As $\beta$ increases beyond $\beta_\mathrm{c}$, more groups are able to self-sustain a local outbreak, until a point where the epidemic delocalizes and invades the whole network.
This analytical description is in line with the work of Ref.~\cite{Cota2018}, where numerical evidence for Griffiths phases was found in a similar setting.

To predict the emergence of this phenomenon, we need to have some better intuition of the behavior of $I_n$ near $\beta_\mathrm{c}$.
Since $\rho \to 0$ near the critical point, we write
\begin{align*}
    I_n = \frac{1}{n} H_n'(1;\beta) \rho  + \mathcal{O}(\rho^2) \;.
\end{align*}
Performing a saddle-point approximation for $H_n'(1;\beta)$, we obtain the following asymptotic behavior for large $n$
\begin{align}\label{eq:asymptotic_Hn}
    H_n'(1;\beta) \sim \begin{dcases}
        \frac{n}{1 - \beta n} & \text{if } \beta < n^{-1} \\
        n^{3/2} \;(\beta  n)^{n} \;e^{-n+1/\beta } & \text{if } \beta \geq n^{-1} \;,
    \end{dcases}
\end{align}
where ``$\sim$" means \textit{asymptotically proportional}.
For $\beta = a n^{-1}$ where $a > 1$ is a constant independent of $n$, this implies that $I_n = \mathcal{O}\left (n^{1/2}e^{bn} \right)$ with \mbox{$b > 0$}. A more formal proof could be made following an argument similar to the one used in Appendix \ref{app:bounds_threshold}.
Therefore, near the epidemic threshold (i.e. $\beta = \beta_\mathrm{c} + \epsilon$ with $\epsilon \ll 1$), we expect the epidemic to be \textit{localized} within any groups of size $n$ for which $\beta > n^{-1}$.
More formally, we say that the epidemic is localized near the epidemic threshold when $I_{n_\mathrm{max}} / I_2 = \mathcal{O}(n_\mathrm{max}^{1/2}e^{b n_\mathrm{max}})$, and we then expect a smeared phase transition, such as the one presented in Fig.~\ref{fig:local}(e).
Conversely, if $I_{n_\mathrm{max}} / I_2 = \mathcal{O}(1)$ near $\beta_\mathrm{c}$, then we say that the epidemic is \textit{delocalized}, and we expect a phase transition similar to the one shown on Fig.~\ref{fig:local}(b).

\section{Mesoscopic localization}
\label{sec:mesoscopic_localization}

In this section, we fully characterize the emergence of \textit{mesoscopic localization}, where the epidemic is localized only within the largest groups near $\beta_\mathrm{c}$ for power-law distributions of membership and group size.
In Sec.~\ref{subsec:localization_regimes}, we derive general asymptotic expressions to distinguish the localization regimes, establishing a partition of the $(\gamma_m,\gamma_n)$ space.
We then investigate in Sec.~\ref{subsec:finite_cutoffs} the effect of finite cut-offs on the localization regimes, and how our results relate to earlier works using the inverse participation ratio.

\subsection{Asymptotic localization regimes}
\label{subsec:localization_regimes}

\begin{figure*}[tb]
\centering
\includegraphics[width=\linewidth]{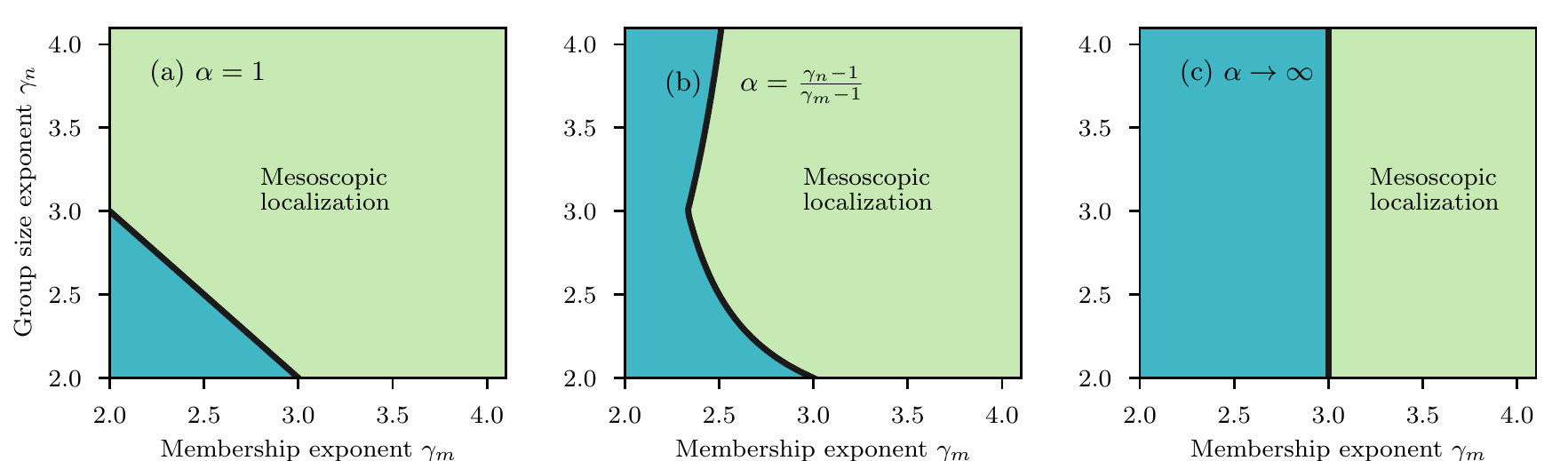}
\caption{Asymptotic localization regimes for power-law membership and group size distributions, and for different cut-off relationships \mbox{$m_\mathrm{max} \sim n_\mathrm{max}^\alpha$}. In the pale green regions, the epidemic is localized near the epidemic threshold $\beta_\mathrm{c}$, while it is delocalized in the darker blue regions. The boundary separating the two regions is inferred from Eqs~(\eqref{eq:threshold_asymptotic_condition}a-c).}
\label{fig:asymptotic_phase_diagrams}
\end{figure*}

Let us assume that both $n_\mathrm{max} \to \infty$ and $m_\mathrm{max} \to \infty$.
As it will be shown, the relation between the cut-offs $n_\mathrm{max}$ and $m_\mathrm{max}$ influences the localization regimes. Henceforth, let us assume a general asymptotic relationship of the form
\begin{align}
    \label{eq:relationship_mmax}
    m_\mathrm{max} \sim n_\mathrm{max}^\alpha \;,
\end{align}
where the exponent $\alpha\geq0$ encodes how both limits $n_\mathrm{max} \to \infty$ and $m_\mathrm{max} \to \infty$ are taken.

To gain some insights on the meaning of Eq.~\eqref{eq:relationship_mmax}, let us assume for the sake of the argument that we have a finite-size network with $N$ nodes and \mbox{$\langle m \rangle N/\langle n \rangle \sim N$} groups.
We could impose cut-offs that are agnostic to the underlying distribution $g_m$ and $p_n$, for instance $m_\mathrm{max} \sim N^{1/2}$ and $n_\mathrm{nmax} \sim N^{1/2}$. This would correspond to $\alpha = 1$. Another option, borrowed from extreme value theory, would be to use the \textit{natural} cut-offs of the two power-law distributions, $m_\mathrm{max} \sim N^{1/(\gamma_m -1)}$ and $n_\mathrm{max} \sim N^{1/(\gamma_n -1)}$ \cite{Boguna2004,Catanzaro2005generation}. This would correspond to \mbox{$\alpha = (\gamma_n - 1)/(\gamma_m -1)$}. Finally, fixing one of the two cut-offs while letting the other go to infinity would correspond to the limit cases $\alpha \to 0$ or $\alpha \to \infty$.

We now turn to the extraction of the asymptotic behavior of the epidemic threshold in the limit $n_\mathrm{max} \to \infty$ for different combinations of $\gamma_n$ and $\gamma_m$---this will inform us on the type of phase transition, i.e. a localized or a delocalized one.

First, we obtain a tighter upper-bound on $\beta_\mathrm{c}$ in the limit $n_\mathrm{max} \to \infty$ for power-law group size distributions $p_n \propto n^{-\gamma_n}$.
Formally, there exists some $n' \in \mathbb{N}$ such that for all $n_\mathrm{max} > n'$,
\begin{align}
    \label{eq:upper_bound_asymptotic}
    \beta_\mathrm{c} \leq \mathrm{min} \left [ \frac{1}{\Omega(g_m,p_n)} \;, \; \frac{1}{n_\mathrm{max}-2} \right ] \;.
\end{align}
Details are provided in Appendix \ref{app:bounds_threshold}, but the general idea is to combine Eq.~\eqref{eq:upper_bound_threshold} with another bound found by forbidding $A_{n_\mathrm{max}}$ to grow exponentially with $n_\mathrm{max}$.
The lower bound of Eq.~\eqref{eq:lower_bound_threshold} and the upper bound of Eq.~\eqref{eq:upper_bound_asymptotic} tightly constrain the asymptotic behavior of $\beta_\mathrm{c}$, which we write as
\begin{align}
    \label{eq:asymptotic_threshold}
    \beta_\mathrm{c}^{-1} \sim \Omega(g_m,p_n) + n_\mathrm{max} \;.
\end{align}

Second, let us examine the asymptotic behavior of the coupling $\Omega(g_m,p_n)$. The first factor in Eq.~\eqref{eq:coupling} has the following behavior
\begin{align}\label{asymptotic_excess_m}
    \frac{\langle m(m-1) \rangle}{\langle m \rangle} &\sim
    \begin{dcases}
        n_\mathrm{max}^{\alpha (3-\gamma_m)} & \text{if } \gamma_m < 3 \;,\\
        \alpha \ln n_\mathrm{max} & \text{if } \gamma_m = 3 \;, \\
        1 & \text{if } \gamma_m > 3 \;,
    \end{dcases}
\end{align}
and the second one has a similar form
\begin{align}\label{asymptotic_excess_n}
    \frac{\langle n(n-1) \rangle}{\langle n \rangle} &\sim
    \begin{dcases}
        n_\mathrm{max}^{3-\gamma_n} & \text{if } \gamma_n < 3 \;,\\
        \ln n_\mathrm{max} & \text{if } \gamma_n = 3 \;, \\
        1 & \text{if } \gamma_n > 3 \;.
    \end{dcases}
\end{align}
Combining Eqs.~\eqref{asymptotic_excess_m} and \eqref{asymptotic_excess_n} for different $\gamma_m$ and $\gamma_n$ leads to different scalings for $\Omega(g_m,p_n)$.

As a result, we find three cases for the scaling of $\beta_\mathrm{c}$ in the limit $n_\mathrm{max} \to \infty$:
\begin{enumerate}
    \item $\Omega(g_m,p_n) n_\mathrm{max}^{-1} \to \infty \implies \beta_\mathrm{c} n_\mathrm{max} \to 0\;$,
    \item $\Omega(g_m,p_n) n_\mathrm{max}^{-1} \to \mathcal{O}(1)  \implies \beta_\mathrm{c} n_\mathrm{max} \to q < 1\;$,
    \item $\Omega(g_m,p_n) n_\mathrm{max}^{-1} \to 0 \implies \beta_\mathrm{c} n_\mathrm{max} \to 1\;$.
\end{enumerate}
This classification allows us to asssociate an asymptotic behavior to each pair $(\gamma_m, \gamma_n)$.
If $\gamma_m \geq 3$, we necessarily have $\beta_\mathrm{c} n_\mathrm{max} \to  1$. Otherwise, if $2 <\gamma_m < 3$ and
\begin{subequations}
\label{eq:threshold_asymptotic_condition}
\begin{align}
    \bullet \; \; 2 &< \gamma_n < 3\text{, then}  \notag\\
    \quad &\beta_\mathrm{c} n_\mathrm{max}  \to 
    \begin{dcases}
        0 & \text{if } 3-\gamma_n + \alpha(3-\gamma_m) > 1 \;, \\
        q < 1  & \text{if } 3-\gamma_n + \alpha(3-\gamma_m) = 1 \;, \\
        1 & \text{if } 3-\gamma_n + \alpha(3-\gamma_m) < 1 \;,
    \end{dcases} \label{eq:threshold_asymptotic_condition_1} \\
    \bullet \;\gamma_n &= 3\text{, then} \notag\\
    \quad &\beta_\mathrm{c} n_\mathrm{max}  \to 
    \begin{dcases}
        0 & \text{if } \alpha(3-\gamma_m) \geq 1 \;, \\
        1 & \text{if } \alpha(3-\gamma_m) < 1 \;,
    \end{dcases} \label{eq:threshold_asymptotic_condition_2} \\
    \bullet \;\gamma_n &> 3\text{, then} \notag\\
    \quad &\beta_\mathrm{c} n_\mathrm{max}  \to 
    \begin{dcases}
        0 & \text{if } \alpha(3-\gamma_m) > 1 \;, \\
        q < 1  & \text{if } \alpha(3-\gamma_m) = 1 \;, \\
        1 & \text{if } \alpha(3-\gamma_m) < 1 \;.
    \end{dcases} \label{eq:threshold_asymptotic_condition_3}
\end{align}
\end{subequations}

Note that the asymptotic behavior \mbox{$\beta_\mathrm{c} n_\mathrm{max} \to q < 1$} never fills an area in the $(\gamma_m,\gamma_n)$ space---it is simply a limiting case.
The two other cases fill the $(\gamma_m,\gamma_n)$ space, and we interpret them as different \textit{localization regimes} using the definitions of Sec.~\ref{subsec:heterogeneous_structure}.
In the region where $\beta_\mathrm{c} n_\mathrm{max} \to 0$, we have $I_{n_\mathrm{max}} / I_2 = \mathcal{O}(1)$ near the epidemic threshold, and the epidemic is delocalized since groups of all sizes are involved.
In the region where $\beta_\mathrm{c} n_\mathrm{max} \to 1$, we have, near the epidemic threshold ($\beta = \beta_\mathrm{c} + \epsilon$ with $\epsilon \ll 1$), $\beta > n_\mathrm{max}^{-1}$, and therefore \mbox{$I_{n_\mathrm{max}} / I_2 = \mathcal{O}(n_\mathrm{max}^{1/2}e^{b n_\mathrm{max}})$}. The epidemic is therefore localized, thriving only in the largest groups.

Equations (\ref{eq:threshold_asymptotic_condition}a-c) are thus used to identify the region where we expect mesoscopic localization, as illustrated in Fig.~\ref{fig:asymptotic_phase_diagrams} for different values of $\alpha$.
One striking observation is the ubiquity of mesoscopic localization: for a large portion of the parameter space, we expect a disease to be localized around the largest groups.
It is worth to recall that the average degree of a node is proportional to $\left \langle n(n-1) \right \rangle$, hence sparse networks correspond only to the upper portion $(\gamma_n > 3)$ of the phase diagrams in Fig.~\ref{fig:asymptotic_phase_diagrams}.

\subsection{Finite cut-offs and mesoscopic inverse participation ratio}
\label{subsec:finite_cutoffs}

\begin{figure}[tb]
\centering
\includegraphics[width=\linewidth]{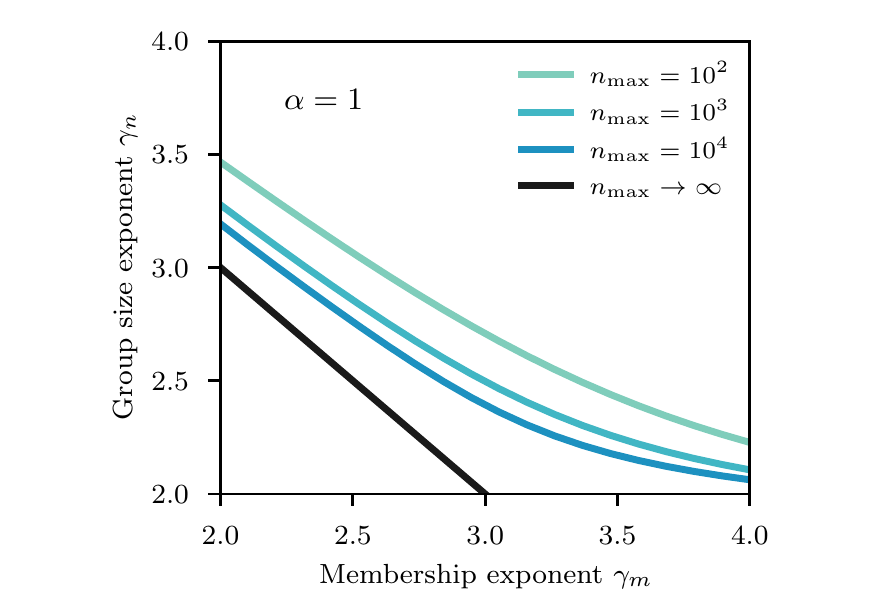}
\caption{Impact of finite cut-offs on the boundary separating the localized and delocalized regimes for power-law membership and group size distributions. We used $m,n \in \lbrace 2,\dots,n_\mathrm{max} \rbrace$, hence $\alpha = 1$, and different values of $n_\mathrm{max}$. Finite cut-offs boundaries are obtained by imposing $\beta_\mathrm{c} = n_\mathrm{max}^{-1}$ and solving Eq.~\eqref{eq:epidemic_threshold} for different pairs $(\gamma_m, \gamma_n)$. The asymptotic case $n_\mathrm{max} \to \infty$ is obtained from Eqs.~(\ref{eq:threshold_asymptotic_condition}a-c).}
\label{fig:finite_size_boundary}
\end{figure}

The results of Sec.~\ref{subsec:localization_regimes} were obtained in the asymptotic limit $n_\mathrm{max} \to \infty$.
However, cut-offs in real systems are always finite.
A finite value for $n_\mathrm{max}$ relaxes the conditions defined in Eqs.~(\ref{eq:threshold_asymptotic_condition}a-c). For a pair $(\gamma_m, \gamma_n)$ in the asymptotically localized regime, it is possible to have either $\beta_\mathrm{c} \gtrless n_\mathrm{max}^{-1}$.
To stay coherent with our definition for a localized epidemic, we must have $\beta_\mathrm{c} \geq n_\mathrm{max}^{-1}$.
Therefore, the condition $\beta_\mathrm{c} \equiv n_\mathrm{max}^{-1}$ leads to the finite cut-offs boundary, given by solutions to Eq.~\eqref{eq:epidemic_threshold} in terms of $\gamma_m$ and $\gamma_n$.
In Fig.~\ref{fig:finite_size_boundary}, we illustrate the boundary separating the delocalized and localized regimes for increasing values of $n_\mathrm{max}$, slowly converging on the asymptotic conditions.
The size of the mesoscopic localization region is smaller compared with that of the asymptotic limit, but it still fills most of the parameter space corresponding to sparse networks ($\gamma_n > 3$).

\begin{figure*}[tb]
\centering
\includegraphics[width=\linewidth]{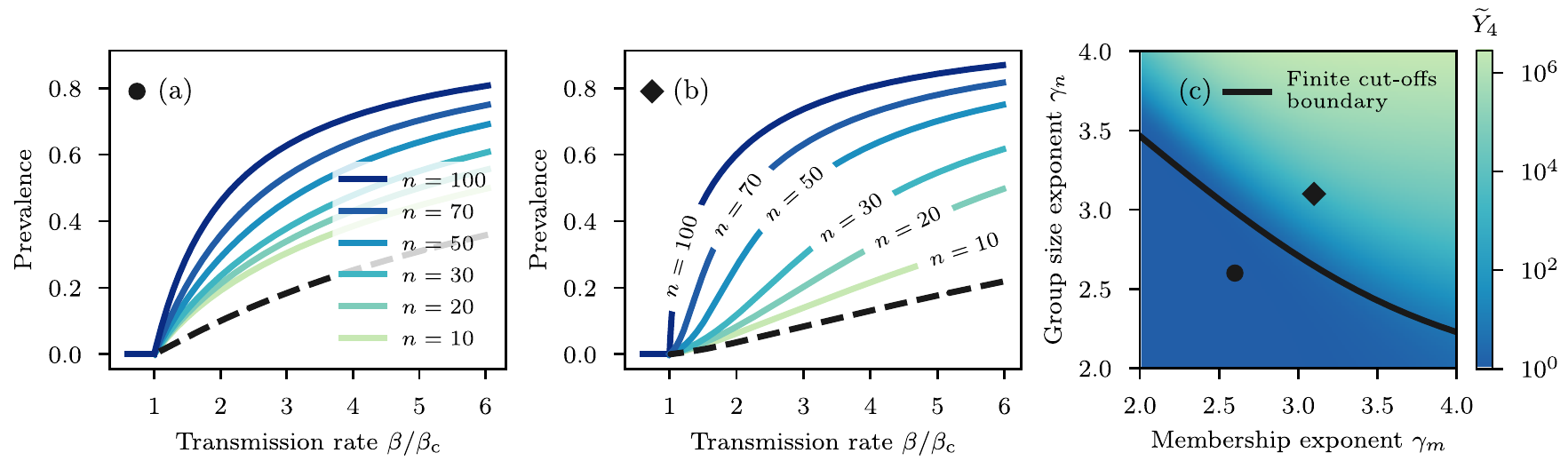}
\caption{Comparison of the stationary state and the level of localization near the epidemic threshold for networks closer to the finite cut-offs boundary. We used power-law membership and group size distributions with $m,n \in \lbrace 2, \dots,100 \rbrace$. (a)-(b) Group prevalence (solid lines) and global prevalence (dashed line) as a function of the transmission rate. (a) $\gamma_m = \gamma_n = 2.6$, yielding $\beta_\mathrm{c} < n_\mathrm{max}^{-1}$. (b) $\gamma_m = \gamma_n = 3.1$, yielding $\beta_\mathrm{c} > n_\mathrm{max}^{-1}$. (c) Quantification of the mesoscopic localization phenomenon using the inverse participation ratio defined at Eq.~\eqref{eq:IPR}.  The solid line corresponds to the boundary between the localized and delocalized regimes, obtained by imposing $\beta_\mathrm{c} = n_\mathrm{max}^{-1}$ and solving Eq.~\eqref{eq:epidemic_threshold} for different pairs $(\gamma_m, \gamma_n)$.}
\label{fig:finite_size_IPR}
\end{figure*}

Another consequence of finite cut-offs is to blur the line between localized and a delocalized epidemic.
Taking pairs $(\gamma_m,\gamma_n)$ closer to the finite-size boundary, we show how this affects the group prevalence in Fig.~\ref{fig:finite_size_IPR}(a) and Fig.~\ref{fig:finite_size_IPR}(b), with $\beta_\mathrm{c} < n_\mathrm{max}^{-1}$ and $\beta_\mathrm{c} > n_\mathrm{max}^{-1}$ respectively.
Near $\beta_\mathrm{c}$, we still associate Fig.~\ref{fig:finite_size_IPR}(a) and \ref{fig:finite_size_IPR}(b) with a delocalized and localized outbreak respectively, but the difference is less marked compared to Fig.~\ref{fig:local}(a) and Fig.~\ref{fig:local}(b).
Therefore, even though the dichotomy is sharp and clear in the asymptotic limit $n_\mathrm{max} \to \infty$, we need to keep in mind that for realistic systems, localization lives on a spectrum. Our next goal is to \textit{quantify} mesoscopic localization.

\textit{At the node level}, an epidemic is considered localized if the contagion is mostly present within a subset of nodes \mbox{$\mathcal{L} \subset \mathcal{V}$}, referred to as the localization set, and \mbox{$\mathcal{V} = \lbrace 1, \dots, N \rbrace$} is the set of all nodes.
An important result from quenched mean-field theory is that the marginal probability for each node $j$ of being infected near $\beta_\mathrm{c}$ is proportional to $v_j$, where $\lbrace v_j \rbrace_{j\in \mathcal{V}}$ are the elements of the principal eigenvector (PEV) of the adjacency matrix.
Epidemic localization can thus be mapped onto eigenvector localization \cite{Martin2014,Pastor-Satorras2016,Castellano2017,Pastor-Satorras2018,Sharkey2019}.
With a normalized eigenvector satisfying $\sum_{j} v_j^2 \equiv 1$, a completely delocalized epidemic at the level of nodes implies \mbox{$v_j \sim N^{-1/2} \; \forall \; j \in \mathcal{V}$}, while a purely localized one corresponds to \mbox{$v_j \sim |\mathcal{L}|^{-1/2} \; \forall \; j \in \mathcal{L}$} and $v_j \sim 0 \;\forall\; j \not\in \mathcal{L}$.
A standard scalar to quantify the localization is the inverse participation ratio $Y_4(N)$. We use the following rescaled version
\begin{align}
    \label{eq:IPR_standard}
    \widetilde{Y}_4(N) &\equiv N \sum_{j = 1}^N v_j^4 \;.
\end{align}
For a delocalized eigenvector, $\widetilde{Y}_4(N) \sim 1$, while for a localization set of size $|\mathcal{L}| \sim N^{\delta}$, then $\widetilde{Y}_4(N) \sim N^{1-\delta}$.
Consequently, $\widetilde{Y}_4^{-1}$ is an effective measure for the fraction of nodes belonging to the localization set.

\textit{At the mesoscopic level}, we consider an epidemic localized if the contagion is mostly present within a subset of the groups.
The difference is subtle, but important : if we observe a delocalized epidemic at the mesoscopic level, it could still be localized at the level of nodes.
To quantify mesoscopic localization, we use an inverse participation ratio as well
\begin{align}
    \label{eq:IPR}
    \widehat{Y}_4(p_n) &= \frac{\sum_n p_n I_n^4}{\left (\sum_n p_n I_n^2 \right )^2} =
    \begin{dcases}
        1 & \text{if } I_n \propto 1 \; \forall n \;,\\
        p_{n'}^{-1} & \text{if } I_n \propto \delta_{n,n'} \;.
    \end{dcases}
\end{align}
As a result, $\widehat{Y}_4^{-1}$ is an effective measure for the fraction of \textit{groups} participating to the epidemic.
Interestingly, Eq.~\eqref{eq:IPR} can be obtained with our analytical formalism, using $I_n$ evaluated at the epidemic threshold $\beta_\mathrm{c}$, or through the connection with quenched mean-field theory.
In the latter case, one extracts the PEV of a network with cliques, then compute
\begin{align}\label{eq:IPR_empirical}
    I_n \propto \frac{1}{|\mathcal{C}_n|}\sum_{\mathcal{S} \in \mathcal{C}_n} \sum_{j \in \mathcal{S}} \frac{v_j}{n} \;,
\end{align}
where $\mathcal{C}_n$ is the set of cliques of size $n$ and $\mathcal{S}$ is the set of nodes belonging to a specific clique. The group distribution then correspond to $p_n \propto |\mathcal{C}_n|$.
Note that this measure relies on an explicit knowledge of $\mathcal{C}_n$, which is already given for synthetic networks (see Appendix \ref{app:network_generation}), or could be extracted using a clique decomposition for real networks.

In Fig.~\ref{fig:finite_size_IPR}(c), we illustrate the behavior of $\widehat{Y}_4(p_n)$ as a function of $(\gamma_m,\gamma_n)$, obtained with our analytical formalism.
As expected, the inverse participation ratio changes drastically near the boundary separating the delocalized and localized regimes for finite cut-offs.
The change would become sharper and sharper as we let $n_\mathrm{max} \to \infty$, and the position of the boundary would move closer to the asymptotic limit, as in Fig.~\ref{fig:finite_size_boundary}.
This inverse participation ratio is therefore a good measure for mesoscopic localization, and could be used to get insights on how the epidemic changes from a localized to a delocalized phase as we increase $\beta$ beyond $\beta_\mathrm{c}$.

\begin{figure}[tb]
\centering
\includegraphics[width=\linewidth]{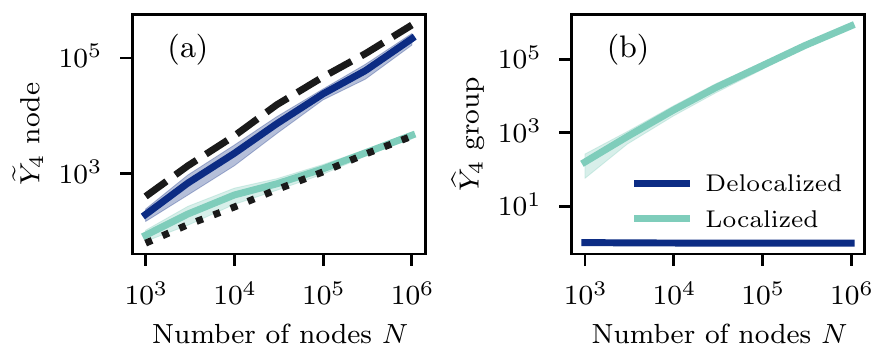}
\caption{Mesoscopic localization is imperceptible using the standard inverse participation ratio on nodes. We performed a finite-size scaling of the inverse participation ratio in the delocalized and mesoscopic localization regime, for nodes and groups. We generated multigraphs of various sizes with different power-law membership and group size distributions [see Appendix \ref{app:network_generation}]. In the delocalized regime, we used $\gamma_m  = 2.3$ and $\gamma_n = 3.5$ ; in the localized regime, we used $\gamma_m = \gamma_n = 3.5$. In both regimes, $m \in \lbrace 2, \dots,m_\mathrm{max} \rbrace$ and $n \in \lbrace 2, \dots,n_\mathrm{max} \rbrace$ with natural cut-offs $m_\mathrm{max} = N^{1/(\gamma_m - 1)}$ and $n_\mathrm{max} = N^{1/(\gamma_n - 1)}$. (a) The solid lines represent the average inverse participation ratio $\widetilde{Y}_4$ for nodes [Eq.~\eqref{eq:IPR_standard}]. The dashed (dotted) line is the average inverse of the fraction of nodes associated to the maximal $K-$core in the delocalized (localized) regime. (b) The solid lines represent the average inverse participation ratio $\widehat{Y}_4$ for groups [Eq.~\eqref{eq:IPR}]. We extracted $I_n$ from the PEV. The shaded regions in both panels correspond to one standard deviation.}
\label{fig:IPR_comparison}
\end{figure}

In Fig.~\ref{fig:IPR_comparison}, we compare the finite-size scaling of the inverse participation ratios for nodes and groups, obtained by generating synthetic networks in the delocalized and localized regime and extracting their PEV.
Although our analytical formalism effectively describes groups of a sub-extensive size, this is not a necessary condition to observe mesoscopic localization.
We have therefore relaxed this assumption to generate the synthetic networks: we have used cut-offs that scale with the number of nodes $m_\mathrm{max} = N^{1/(\gamma_m - 1)}$ and $n_\mathrm{max} = N^{1/(\gamma_n - 1)}$. These are more appropriate for the finite-size scaling analysis.

In Fig.~\ref{fig:IPR_comparison}(a), we see that the inverse participation ratio for nodes $\widetilde{Y}_4$ increases in both the delocalized and the mesoscopic localization regime.
It scales similarly to the inverse fraction of the nodes belonging to the maximal $K-$core, in agreement with previous works on the subject \cite{Castellano2012,Castellano2017,Pastor-Satorras2018}. The localization set can thus be associated with the innermost core in both cases, and despite a different scaling law, there is no clear sign of a change of regime between the two curves.
Figure~\ref{fig:IPR_comparison}(b) tells us another story: the inverse participation ratio for groups $\widehat{Y}_4$ converges to $1$ in the delocalized regime, but scales as a power law in the mesoscopic localization regime, clearly indicating a transition of regime.

Figure~\ref{fig:IPR_comparison} strongly advocates for a change of perspective if we want to detect potentially hidden localized phase at the mesoscopic level. We need to focus on the higher-level organization, the groups, and find better ways to characterize their impact on the dynamics. If we focus our attention at the node level, Fig.~\ref{fig:IPR_comparison}(a) tells us that an epidemic localized at the mesoscopic level is no different from a delocalized one---the contagion is mostly present within the innermost core in both cases. However, the composition of this core and of the outer shells is quite different, as can be inferred from $\widehat{Y}_4$ in Fig.~\ref{fig:IPR_comparison}(b). In the localized regime, the innermost core is composed mostly of the largest groups, while groups of all sizes compose the core in the delocalized regime. Recall that a bias toward larger groups has dramatic consequences on the dynamics, leading to a smeared phase transition instead of a clean one.

Before closing this section, it is probably useful to stress once more the versatility and generality of our approach. The results on synthetic networks are representative of results that can also be obtained on real complex systems. More complex networks are generally made of mesoscopic substructures, dense or sparse---not necessarily cliques---with possibly weighted and directed edges. The important observation is that the details of these substructures do not matter much. As long as it is possible to identify them, using community detection \cite{Fortunato2010,Fortunato2016}, random clique cover \cite{williamson2020}, or hypergraph reconstruction \cite{young2020}, one can construct a higher-order representation of the original network with nodes belonging to groups and evaluate the localization on these groups using a measure similar to Eq.~\eqref{eq:IPR_empirical}. Following this line, the original structure does not even need to be a network of pairwise interactions. It could already be a higher-order representation, such as a simplicial complex or a hypergraph \cite{Battiston2020}.

\section{Discussion}
\label{sec:discussion}

One of the important factors behind the success of network science to study contagions, from infectious diseases to the spread of information, is that it provides a mathematical framework to go beyond the assumption of a homogeneous population \cite{Pastor-Satorras2015}.
Contagions are rarely driven by the average individual, mostly because some individuals are simply more connected than others but also potentially more central.
Beyond the fact that they drive the dynamics of contagions, these key actors are also critical to their control.
On the one hand, it allows the mathematical formulation of targeted immunization and interventions \cite{pastor2002immunization, hunter2019social}: Which individuals should be immunized or removed from the network to minimize the spread on an infectious disease?
On the other hand, it also permits the identification of influential spreaders \cite{morone2015influence}: Which individuals should seed a contagion in order to maximize its spread?
These different ideas all revolve around a control theory for contagions, but also all depend on a theoretical understanding of what type of structures matters for contagions.

In practice, however, social networks are not randomly mixed but contain a higher-level organization determined by workplaces, schools, events, etc.; such that key actors can be places, social gatherings, or more abstractly groups, rather than the individuals themselves.
Thankfully, multiple new approaches to handle higher-order interactions have been proposed in recent years.
In the thermodynamic limit, the networks used in this paper can equivalently be represented using ideas of topological simplexes from topology \cite{iacopini2019simplicial}, hypergraphs \cite{jhun2019simplicial, de2019social}, or projections of bipartite networks \cite{Newman2003, hebert2010propagation}.
Under the right level of mean-field approximation, these are all equivalent.
However, their dynamics at the mesocopic level can be very heterogeneous, as in Fig.~\ref{fig:local}(f), since groups can take considerably more different states than individuals who are usually only susceptible or infected.
Therefore, adequate care should be exercised not to over-simplify (coarse grain) the mathematical description in order to embrace this heterogeneity. Our group-based approximate master equation framework acknowledges fully this warning.

Using this approach, we have observed and analyzed a phenomenon of mesoscopic localization where contagions can concentrate around groups that are large enough to allow a local, self-sustained outbreak in the presence of some weak external group coupling.
Interestingly, while there is little empirical evidence for localization of real contagions around hubs in a contact network, there are well-known cases of dynamics resembling mesoscopic localization.
For example, bacterial infections in hospitals (e.g. \textit{C. difficile} \cite{mcfarland1986review}) are already a well-documented example of mesocopically localized contagions, but are simply never studied analytically as such.

In this mesocopic localization phase, influential groups are naturally found to be the larger ones around which a contagion can localize.
Intervention or control operating at a structural level (i.e. on groups rather than on individuals) should therefore focus around these influential groups.
The large toolbox developed for targeted immunization \cite{hebert2013global} and identification of influential spreaders \cite{weng2014predicting} could now be leveraged, at the mesoscopic level, to better understand and control contagions on networks capable of mesocopic localization.
In Ref.~\cite{st2020school}, we investigate the impact of removing groups as a model of school closures and event cancellations.
We find that delocalized dynamics are characterized by a linear relationship between outbreak size and the strength of our intervention, akin to mass-action models. Conversely, localized dynamics show a non-linear relationship that varies with the importance of the localization effects. For strongly localized epidemics, there is an increasing effectiveness of interventions, leading to a sudden collapse of the epidemic.

In a broader context, higher-order structures were found to be important for a wide range of dynamics, from competitive dynamics \cite{grilli2017higher} to social contagion \cite{iacopini2019simplicial}.
Several of these studies highlight non-trivial effects of higher-order structures on dynamics using numerical tools or very coarse-grained analytical methods.
These approaches, ignoring the heterogeneous states of groups, limit the type of questions and behaviors that can be answered and analyzed.
We wish to emphasize that master equation descriptions provide valuable insights into the mechanisms of these dynamics and their interplay with higher-order structures. For instance, we conjecture that mesoscopic localization is even more present in systems with social reinforcement mechanisms \cite{osullivan2015mathematical}, and that its impacts on the global state of the dynamics are even more dramatic.

There are now several avenues open to broaden the applicability of our simple approach.
In its current form, the only inputs required are a membership distribution $g_m$ and a group size distribution $p_n$, along with the specification of the local dynamics.
As a first step, our future works will focus on improving the heterogeneous mean-field coupling between master equations.
We could, for example, refine our description of the states of the nodes in order to capture dynamical correlations with the state of the groups in which they are found, include structural correlations between the memberships of nodes and the sizes of groups through a joint distributions $P(m,n)$, or allow groups with more complex inner contact patterns.
We hope that our work on mesoscopic localization and the framework that has emerged will provide a solid foundation for the continuing efforts to improve our understanding of dynamics on complex networks.

\section*{Acknowledgments}
The authors acknowledge Calcul Qu\'{e}bec for computing facilities.
This work was supported by the National Institutes of Health 1P20 GM125498-01 Centers of Biomedical Research Excellence Award (L.H.-D.), the Fonds de recherche du Qu\'{e}bec – Nature et technologies (V.T., G.S.), the Natural Sciences and Engineering Research Council of Canada (G.S., V.T., A.A., L.J.D.), and the Sentinelle Nord program of Universit\'{e} Laval, funded by the Canada First Research Excellence Fund (G.S., V.T., A.A., L.J.D.).

\appendix

\section{Detailed development for the epidemic threshold}
\label{app:epidemic_dev}

The function $F(\rho)$ corresponds to
\begin{align*}
    F(\rho) = r(\rho)  \left [ \frac{\sum_m m(m-1) s_m(\rho) g_m}{ \sum_m m s_m (\rho) g_m} \right ] \;.
\end{align*}
To find its derivative with respect to $\rho$, let us note that \mbox{$s_m(\rho) = 1 + \mathcal{O}(\rho)$} and $r(\rho) = \mathcal{O}(\rho)$ as $\rho \to 0$, which can be deduced from Eq.~\eqref{eq:mean-field1} by using $c_{n,i} = \delta_{i,0} + h_{n,i} \rho + \mathcal{O}(\rho^2)$.
Therefore,
\begin{align}\label{eq:appendix_A_F_derivative}
    \left. \frac{\mathrm{d} F}{\mathrm{d}\rho} \right |_{\rho \to 0} &= \frac{\langle m(m-1) \rangle}{\langle m \rangle} \left. \frac{\mathrm{d} r}{\mathrm{d}\rho} \right |_{\rho \to 0} \;,
\end{align}
and the derivative of $r(\rho)$ is
\begin{align}\label{eq:appendix_A_r_derivative}
    \left.  \frac{\mathrm{d}r}{\mathrm{d}\rho} \right |_{\rho \to 0} &= \frac{ \sum_{n,i} \beta i(n-i) h_{n,i} p_n}{\sum_{n} n  p_n} \;.
\end{align}
In terms of the generating functions $H_n(x;\beta)$, with
\begin{align*}
    H_n'(1;\beta) &= \sum_{i} i h_{n,i} \;, \\
    H_n''(1;\beta) &= \sum_{i} i(i-1) h_{n,i} \;,
\end{align*}
we obtain the relation,
\begin{align}\label{eq:appendix_A_relation}
    \sum_i i(n-i) h_{n,i} &=  (n-1) H_n'(1;\beta) - H_n''(1;\beta) \;.
\end{align}
Combining Eqs.~\eqref{eq:appendix_A_F_derivative}, \eqref{eq:appendix_A_r_derivative}, and \eqref{eq:appendix_A_relation}, we arrive at
\begin{align}\label{eq:rho_derivative}
    \left. \frac{\mathrm{d} F}{\mathrm{d}\rho} \right |_{\rho \to 0} &= \beta \frac{\langle m(m-1) \rangle}{\langle m \rangle \langle n \rangle} \left \langle  (n-1) H_n'(1;\beta) - H_n''(1;\beta) \right \rangle  \;.
\end{align}

With the integral representation of $H_n(x;\beta)$, Eq.~\eqref{eq:gen_func_Hn}, and the derivatives
\begin{align*}
    H_n'(1;\beta) &= n \int_0^\infty (1+\beta u)^{n-1} e^{-u} \mathrm{d} u \;, \\
    H_n''(1;\beta) &=  n(n-1) \int_0^\infty (1+\beta u)^{n-2} \beta u e^{-u} \mathrm{d} u \;,
\end{align*}
we end up with the simplification
\begin{align*}
    (n-1)H_n'(1;\beta) - H_n''(1;\beta) &= n(n-1)A_n(\beta) \;,
\end{align*}
where
\begin{align}
    \label{eq:An_appendix}
    A_n(\beta) \equiv \int_0^\infty (1+\beta u)^{n-2} e^{-u} \mathrm{d} u \;.
\end{align}
Inserting in Eq.~\eqref{eq:rho_derivative} and setting the derivative to 1, we finally obtain the implicit expression for the epidemic threshold given by Eq.~\eqref{eq:epidemic_threshold}.

\section{Bounds on the epidemic threshold}
\label{app:bounds_threshold}

Let us bound $\beta_\mathrm{c}$ for any $n_\mathrm{max}$ by bounding $A_n(\beta)$ [Eq.~\eqref{eq:An_appendix}] for all $n$.
First, since $\beta u \geq 0$, then
\begin{align}
    \label{eq:lower_bound_An}
    A_n(\beta) \geq \int_0^\infty e^{-u} \mathrm{d} u = 1 \;.
\end{align}
Second, we rewrite
\begin{align*}
    A_n(\beta) = \int_0^\infty e^{\phi(u ; \beta)} \mathrm{d} u \;,
\end{align*}
where $\phi(u;\beta) = (n-2)\ln ( 1 + \beta u) - u$. Since $\ln(1+x) \leq x$, $\phi(u ; \beta) \leq \beta (n-2) u - u$, which implies
\begin{align*}
    A_n(\beta) \leq
    \begin{dcases}
        \infty & \text{if } \beta (n-2) \geq 1 \;, \\
        \frac{1}{1 - \beta (n-2)} & \text{if } \beta (n-2) < 1 \;.
    \end{dcases}
\end{align*}
We relax the conditions by replacing $n$ by $n_\mathrm{max}$ everywhere on the right-hand side
\begin{align}
    \label{eq:upper_bound_An}
    A_n(\beta) \leq
    \begin{dcases}
        \infty & \text{if } \beta (n_\mathrm{max}-2) \geq 1 \;, \\
        \frac{1}{1 - \beta (n_\mathrm{max}-2)} & \text{if } \beta (n_\mathrm{max}-2) < 1 \;.
    \end{dcases}
\end{align}
By inserting Eqs.~\eqref{eq:lower_bound_An} and \eqref{eq:upper_bound_An} in Eq.~\eqref{eq:epidemic_threshold} and solving for $\beta$, we find the bounds of Eq.~\eqref{eq:bounds_threshold}. Only the second case of Eq.~\eqref{eq:upper_bound_An} leads to a coherent bound for $\beta_\mathrm{c}$.

The upper bound on the epidemic threshold is not very tight, but we can do better if we assume $p_n \propto n^{-\gamma_n}$ and the limit \mbox{$n_\mathrm{max} \to \infty$}.
It follows that there exists some $n' \in \mathbb{N}$ such that for all $n_\mathrm{max} > n'$, the epidemic threshold must respect \mbox{$\beta_\mathrm{c} \leq (n_\mathrm{max} - 2)^{-1}$}.
Let us make a proof by contradiction: we start with the premise that \mbox{$\beta = a (n_\mathrm{max} - 2)^{-1}$} for some arbitrary constant $a > 1$.
We know that \mbox{$\ln(1+x) \geq x(1-x)$} for all $x \geq 0$, hence
\begin{align*}
    \phi(u ; \beta) \geq -\beta^2(n-2)  u^2 + [\beta(n-2) -1]u \;.
\end{align*}
Making the change of variable $y = \beta \sqrt{n-2} u$ and defining $d \equiv [\beta(n-2) -1]/(2\beta\sqrt{n-2})$, we arrive at
\begin{align*}
    A_n(\beta) &\geq \frac{e^{d^2}}{\beta \sqrt{n-2}} \int_0^\infty e^{-(y-d)^2} \mathrm{d} y \quad \forall n > 2\;,\\
        &\geq \frac{\sqrt{\pi} e^{d^2}}{2\beta \sqrt{n-2}} \;.
\end{align*}
Let us focus on $n = n_\mathrm{max}$. In this case, using our premise for $\beta$, we have
\begin{align*}
    d^2 = \frac{(a-1)^2(n_\mathrm{max}-2)}{4 a^2} \equiv  b (n_\mathrm{max}-2) \;,
\end{align*}
where $b>0$. Therefore, there always exists a constant $B_1 > 0$ independent from $n_\mathrm{max}$ and $a$ such that
\begin{align*}
A_{n_\mathrm{max}}(\beta) \geq  \frac{1}{a}\left( \frac{1}{2}\sqrt{\pi(n_\mathrm{max}-2)}e^{-2b}\right) e^{b n_\mathrm{max}}  \geq \frac{B_1}{a} e^{b n_\mathrm{max}} \;,
\end{align*}
This provides a lower bound for the following term
\begin{align*}
    \left \langle n(n-1) A_n(\beta) \right \rangle &\geq p_{n_\mathrm{max}} n_\mathrm{max}(n_\mathrm{max}-1) A_{n_\mathrm{max}}(\beta) \;, \\
                                                   &\geq \frac{B_2}{a} n_\mathrm{max}^{2 - \gamma_n} e^{b n_\mathrm{max}} \;,
\end{align*}
where we assumed \mbox{$p_n \propto n^{-\gamma_n}$} with $\gamma_n < \infty$.
For some constant $B_2 < \infty$.
Inserting this and our premise on $\beta$ in Eq.~\eqref{eq:epidemic_threshold}, we obtain an expression of the form
\begin{align}
    \label{eq:contradiction}
    n_\mathrm{max}^{1 - \gamma_n} \leq B_3  e^{-b n_\mathrm{max}} \;.
\end{align}
For some constant $B_3 < \infty$. Equation \eqref{eq:contradiction} is clearly not respected in the limit $n_\mathrm{max} \to \infty$, hence completing the proof by contradiction.

Note that a solution $\beta_\mathrm{c} > (n_\mathrm{max} - 2)^{-1}$ is not ruled out if $p_n$ decrease exponentially for large $n$.

\section{Generation of networks with cliques}
\label{app:network_generation}

We generated multigraphs using a stub matching process.
First, each node $j \in \mathcal{V}$ is assigned a membership $m$ drawn from $g_m$, resulting in a membership sequence \mbox{$\bm{m} = (m_1, m_2, \dots, m_N)$}.
Then, we create a group size sequence of length $N'$, $\bm{n} = (n_1, n_2, \dots, n_{N'})$, by drawing sizes $n_k$ according to $p_n$.
We must additionally constrain the sequence such that the number of membership stubs and the number of group stubs (available spot for the nodes) are the same
\begin{align}\label{eq:stub_constraint}
    \sum_{j=1}^N m_j = \sum_{k=1}^{N'} n_k \;.
\end{align}
In practice, if the right-hand side of Eq.~\eqref{eq:stub_constraint} is smaller than the left-hand side, we add another group with size $n$ drawn from $p_n$.
If it is bigger, we remove a group uniformly at random.
We repeat this process until the number of stubs is equal on both sides.
$N'$ is therefore not fixed, but it is expected that $N' \sim \langle m\rangle N/\langle n\rangle$ since both $\langle m\rangle$ and $\langle n\rangle$ are bounded.

Once we have the membership and group size sequences, we match the stubs uniformly at random---an edge is added between each pair of nodes belonging to a same group.
This effectively creates loopy multigraphs, but the loops and multi-edges represent a vanishing fraction of the total number of edges for $N \to \infty$ ; we do not remove them since they have a marginal impact on the dynamics.

\section{Validation with Monte Carlo simulations}
\label{app:MC_validation}

\begin{figure}[tb]
\centering
\includegraphics[width=\linewidth]{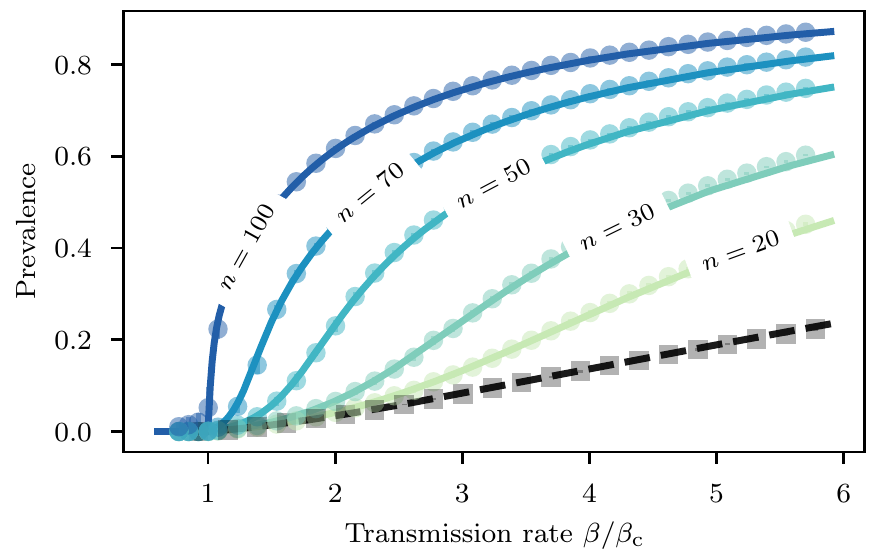}
\caption{Validation of the approximate master equation approach with Monte Carlo simulations, using the efficient algorithm provided by Ref.~\cite{St-Onge2019}. We used a homogeneous membership distribution $g_m = \delta_{m,3}$, a heterogeneous group size distribution of the form $p_n \propto n^{-\gamma_n}$, and $n \in \lbrace 2,\dots,100 \rbrace$. The solid lines represent the group prevalence and the dashed line represents the global prevalence obtained with Eqs.~\eqref{eq:stationary_state_sm} and \eqref{eq:cni_explicit}. The markers represent the average group (circles) and global (squares) prevalences in the quasi-stationary state over 20 network realizations of size $N = 5\times 10^5$. The error bars (smaller than the markers) represent the standard deviation over these 20 realizations. For each network, 20\% of the nodes were initially infected at random, then after a burn-in period $\Delta t \in [50,1000]$, between 100 and 1000 states have been sampled to estimate the prevalences, each separated by a decorrelation period $\Delta t \in [1,10]$. Larger values for the burn-in period, the decorrelation period and the number of states sampled were used near the epidemic threshold. The epidemic threshold $\beta_\mathrm{c}$ is the solution to Eq.~\eqref{eq:epidemic_threshold}.
}
\label{fig:mc_validation}
\end{figure}

In Fig.~\ref{fig:mc_validation}, we compare the predictions of our approximate master equation approach with the results of extensive Monte Carlo simulations.
Our analytical framework accurately reproduce the behavior of the SIS model on synthetic networks generated with the method presented in Appendix~\ref{app:network_generation}.
Figure \ref{fig:mc_validation} also confirms the existence of the mesoscopic localization phenomenon predicted by our approach.

To simulate the SIS model on multigraphs, we used the efficient algorithm provided by Ref.~\cite{St-Onge2019} to evolve the state of the system.
Since the system typically reaches the absorbing-state for finite-size networks near the epidemic threshold, we sampled the \textit{quasi-stationary state} using the state-of-the-art method presented in Refs.~\cite{deoliveira2005,sander2016}. We kept a history of 100 previous states that were each updated at a rate \mbox{$\omega \in [10^{-3},10^{-2}]$} by the current state of the system. If the system fell on the absorbing-state, it was replaced by a random state picked in the history---after a sufficient burn-in period, this method samples the quasi-stationary state.

\section{Scaling the transmission rate with group size}
\label{app:scaling_transmission}

In the approximate master equations \eqref{eq:ame_ode}, an infected node in a group transmits the disease to all susceptible nodes at rate $\beta$.
Even though it is reasonable to have more infections within large groups, an individual might not interact with all others as much as within smaller groups.
For instance, assume two groups of size $n$ and $n'$, with $n > n'$, the first representing a workplace and the second a household.
An infected individual belonging to both interact with more people in the first, but the strength of the interaction is more important in the second.

Fortunately, our framework is highly flexible.
We could replace the term $\beta i$ in Eq.~\eqref{eq:ame_ode_cni} by a general infection function $f(n,i)$ for the nodes in the group.
For the matter at hand, we simply scale the transmission rate as $\beta \mapsto \beta n^{-\nu}$ with $0 \leq \nu \leq 1$, assuming that, on average, the interaction strength decreases with the group size.
The analysis already performed for $\nu = 0$ is extended to arbitrary values of $\nu$ by direct substitutions.

\subsection{Epidemic threshold}
\label{app:epidemic_threshold}

First, we have the following new definition for $r(\rho)$ in the stationary state
\begin{align*}
    r(\rho) &= \frac{  \sum_{n,i} \beta n^{-\nu} i(n-i)c_{n,i}(\rho) p_n}{\sum_{n,i} (n-i) c_{n,i}(\rho) p_n} \;.
\end{align*}
Near the absorbing-state, we redefine the generating function as
\begin{align*}
    H_n(x;\beta,\nu) &= h_{n,0} + \frac{n^\nu}{\beta} \int_{0}^\infty \left [ \left ( 1 + \frac{\beta u x }{n^\nu} \right )^n - 1 \right ] u^{-1} e^{-u} \mathrm{d} u \;.
\end{align*}
The condition for the epidemic threshold then becomes
\begin{align*}
    \left. \frac{\mathrm{d} F}{\mathrm{d} \rho}  \right|_{\rho \to 0} =& \; \beta \frac{\langle m(m-1) \rangle}{\langle m \rangle  \langle n \rangle } \\
        &\times \left \langle n^{-\nu} \left [ (n-1) H_n'(1;\beta,\nu) -  H_n''(1;\beta,\nu) \right ] \right \rangle \;, \\
        \equiv & \;1 \;.
\end{align*}
After some algebraic manipulations, we arrive at a new implicit expression for $\beta_\mathrm{c}$,
\begin{align}
    \label{eq:epidemic_threshold_new}
    \beta \frac{\langle m(m-1) \rangle}{\langle m \rangle  \langle n \rangle } \left \langle n^{1-\nu}(n-1) A_n(\beta,\nu) \right \rangle = 1 \;,
\end{align}
where
\begin{align*}
    A_n(\beta,\nu) \equiv \int_0^\infty \left ( 1+\frac{\beta u}{n^\nu} \right)^{n-2} e^{-u} \mathrm{d} u \;.
\end{align*}

Since $A_n(\beta,\nu)$ has a similar form as Eq.~\eqref{eq:An_appendix}, it is straightforward to reproduce the results of Appendix \ref{app:bounds_threshold} in this more general context. For a power-law distribution $p_n \sim n^{-\gamma_n}$, we have the following asymptotic behavior for the epidemic threshold
\begin{align}\label{eq:asymptotic_threshold_new}
    \beta_\mathrm{c}^{-1} \sim \Omega(g_m,p_n ; \nu) + n_\mathrm{max}^{1-\nu} \;,
\end{align}
where the \textit{coupling} between groups is
\begin{align}
    \label{eq:coupling_new}
    \Omega(g_m,p_n ; \nu) =  \left (\frac{\langle m(m-1) \rangle}{\langle m \rangle} \right) \left ( \frac{ \left \langle n^{1-\nu}(n-1) \right \rangle}{ \langle n \rangle} \right) \;.
\end{align}

\subsection{Behavior near the absorbing-state}

The group prevalence $I_n$ near the absorbing-state can be estimated from a saddle-point approximation of $H_n'(1;\beta,\nu)$ as well. For large $n$,
\begin{align}\label{eq:asymptotic_Hn_new}
    H_n'(1;\beta,\nu) \sim \begin{dcases}
        \frac{n}{1 - \beta n^{1-\nu}} & \text{if } \beta < n^{\nu-1} \\
        n^{3/2} \;(\beta  n^{1-\nu})^{n} \;e^{-n+n^\nu/\beta } & \text{if } \beta \geq n^{\nu-1} \;.
    \end{dcases}
\end{align}
For $\beta = a n^{\nu-1}$ where $a > 1$ is a constant independent of $n$, we still have $I_n = \mathcal{O}\left (n^{1/2}e^{bn} \right)$ with \mbox{$b > 0$}.
Therefore, $\nu$ affects the value of $\beta$ for which a group of size $n$ can sustain an epidemic locally, but the behavior of $I_n$ is unaltered compared to the $\nu = 0$ case.

\subsection{Mesoscopic localization}

The form of Eqs.~\eqref{eq:asymptotic_threshold_new} and \eqref{eq:coupling_new} is similar to Eqs.~\eqref{eq:asymptotic_threshold} and \eqref{eq:coupling}.
It is then straightforward to obtain the asymptotic localization regimes as in Sec.~\ref{subsec:localization_regimes} by investigating the behavior of $\beta_\mathrm{c} n_\mathrm{max}^{1-\nu}$.
Note that the scaling for the second term of the coupling is now
\begin{align}\label{asymptotic_excess_n_new}
    \frac{\left \langle n^{1-\nu} (n-1) \right \rangle}{\langle n \rangle} &\sim
    \begin{dcases}
        n_\mathrm{max}^{3-\gamma_n-\nu} & \text{if } \gamma_n + \nu < 3 \;,\\
        \ln n_\mathrm{max} & \text{if } \gamma_n + \nu= 3 \;, \\
        1 & \text{if } \gamma_n + \nu > 3 \;.
    \end{dcases}
\end{align}

If $\nu = 1$, we always have $\beta_\mathrm{c} n_\mathrm{max}^{1-\nu} \to q < 1 \text{ or } 0$, hence the epidemic is always \textit{delocalized}.
Therefore, let us focus on $\nu \in [0,1)$. If $\gamma_m \geq 3$, we have $\beta_\mathrm{c} n_\mathrm{max}^{1-\nu} \to 1$ for all $\gamma_n$, as in the case $\nu = 0$, meaning that the outbreak is always \textit{localized}. This is surprising, since $\nu > 0$ increases the value of $\beta$ for which a group of size $n$ is able to sustain an epidemic locally. The reason is that $\nu > 0$ also \textit{decreases} the coupling between groups $\Omega(p_n,g_m; \nu)$, hence both effects cancel each other.
\\

If $ 2 < \gamma_m < 3$  and
\begin{subequations}
\vspace{-5pt}
\label{eq:threshold_asymptotic_condition_new}
\begin{align}
    \quad \bullet \; &2 < \gamma_n + \nu < 3\text{, then}  \notag\\
     & \quad \beta_\mathrm{c} n_\mathrm{max}^{1-\nu}  \to
    \begin{dcases}
        0 & \text{if } 3-\gamma_n + \alpha(3-\gamma_m) > 1 \;, \\
        q < 1  & \text{if } 3-\gamma_n + \alpha(3-\gamma_m) = 1 \;, \\
        1 & \text{if } 3-\gamma_n + \alpha(3-\gamma_m) < 1 \;,
    \end{dcases} \label{eq:threshold_asymptotic_condition_new_1} \\
    \quad \bullet \;&\gamma_n + \nu = 3\text{, then} \notag\\
     & \quad \beta_\mathrm{c} n_\mathrm{max}^{1-\nu}  \to
    \begin{dcases}
        0 & \text{if } \alpha(3-\gamma_m) + \nu \geq 1 \;, \\
        1 & \text{if } \alpha(3-\gamma_m) +\nu  < 1 \;,
    \end{dcases} \label{eq:threshold_asymptotic_condition_new_2} \\
    \quad \bullet \;&\gamma_n + \nu > 3\text{, then} \notag\\
     & \quad \beta_\mathrm{c} n_\mathrm{max}^{1-\nu}  \to
    \begin{dcases}
        0 & \text{if } \alpha(3-\gamma_m) + \nu> 1 \;, \\
        q < 1  & \text{if } \alpha(3-\gamma_m) + \nu  = 1 \;, \\
        1 & \text{if } \alpha(3-\gamma_m) + \nu < 1 \;.
    \end{dcases} \label{eq:threshold_asymptotic_condition_new_3}
\end{align}
\end{subequations}
Again, we see that for $2 < \gamma_n + \nu < 3$, Eq.~\eqref{eq:threshold_asymptotic_condition_new_1}, scaling the transmission rate with $n^{-\nu}$ does not affect the localization regime in the $(\gamma_m,\gamma_n)$ space. The effect becomes perceptible for $\gamma_n + \nu \geq 3$, when the coupling $\Omega(p_n,g_m; \nu)$ is dominated by the first term depending solely on the membership distribution.

\begin{figure}[tb]
\centering
\includegraphics[width=\linewidth]{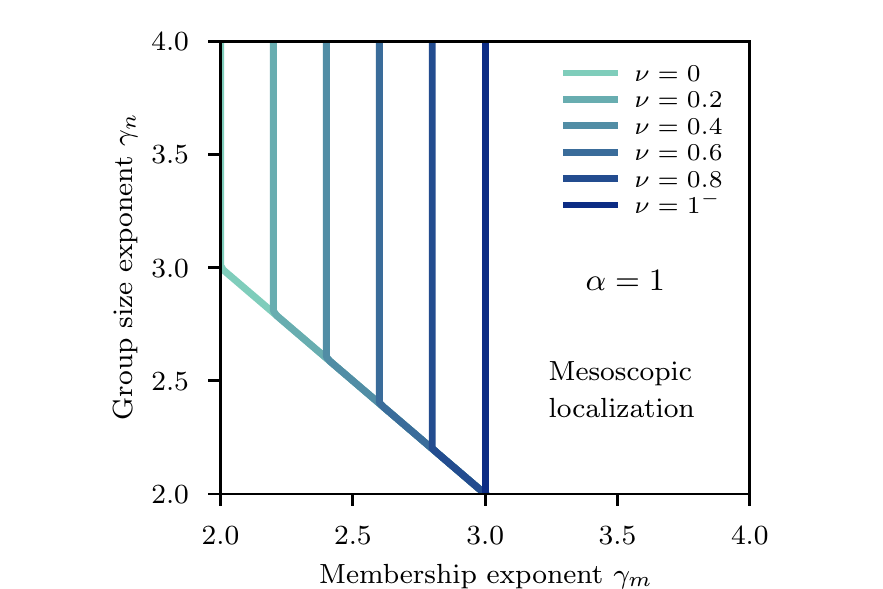}
\caption{Impact of scaling the transmission rate on the boundary separating the localized and delocalized regimes for power-law membership and group size distributions. We used $m,n \in \lbrace 2,\dots,n_\mathrm{max} \rbrace$, hence $\alpha = 1$, and different values of $\nu$ for the relation $\beta \mapsto \beta n^{-\nu}$. The boundaries are obtained from Eqs~(\ref{eq:threshold_asymptotic_condition_new}a-c).}
\label{fig:scaling_boundary}
\end{figure}

Figure~\ref{fig:scaling_boundary} shows the impact of $\nu > 0$ on the boundary separating the localized and delocalized regimes. The top portion of the boundary moves to higher values of $\gamma_m$ as $\nu$ is increased, reducing the size of the mesocopic localization region. In the limit $\nu \to 1^-$, there still exists a non-vanishing portion of the parameter space allowing localization, i.e. $\gamma_m \geq 3$. At $\nu = 1$, mesocopic localization is impossible for all $\gamma_m,\gamma_n$, and thus there is no boundary.

\begin{figure}[tb]
\centering
\includegraphics[width=\linewidth]{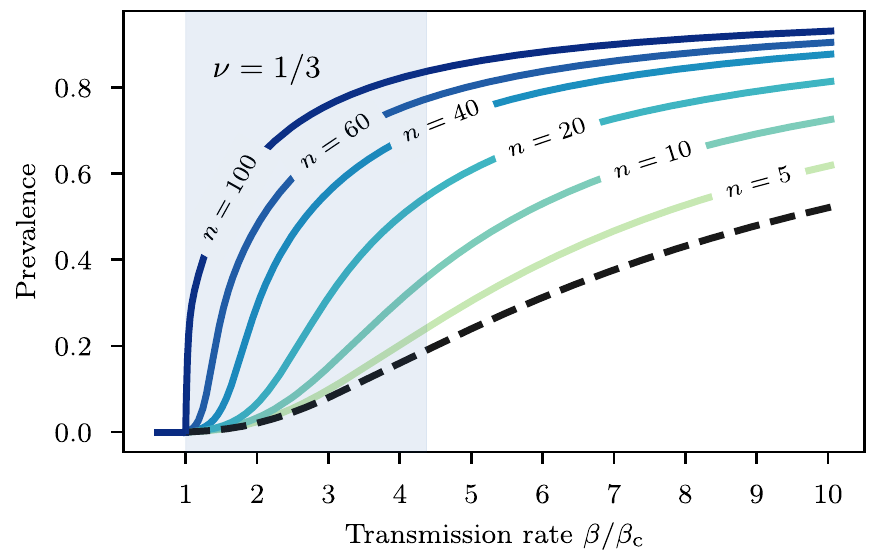}
\caption{Localized portion of the bifurcation diagram in the mesoscopic localization regime. The shaded region represents the localized portion, defined as $[\beta_\mathrm{c}, \Omega(g_m,p_n;\nu)^{-1}]$. We used the same structure as in Fig.~\ref{fig:local}(e), but with $\beta \mapsto \beta n^{-\nu}$, where $\nu = 1/3$. The solid lines represent the group prevalence and the dashed line represents the global prevalence. Stationary state solutions were obtained from Eqs.~\eqref{eq:stationary_state_sm} and \eqref{eq:cni_explicit}. The epidemic threshold $\beta_\mathrm{c}$ is the solution to Eq.~\eqref{eq:epidemic_threshold_new}.
}
\label{fig:local_scaling}
\end{figure}

\section{Localized portion of the bifurcation diagram}

As $\beta$ is increased beyond $\beta_\mathrm{c}$, groups of smaller sizes can self-sustain the epidemic locally, until a point where the disease is present in all groups---the epidemic is not localized anymore.
When the epidemic becomes delocalized, the global prevalence curve reaches an inflexion point---the second derivative with respect to $\beta$ turns negative because all groups sustain the epidemic and saturation effects becomes more important.
This can be seen for instance in Fig.~\ref{fig:local_scaling} around $\beta/\beta_\mathrm{c} = 4.3$.
But how do we define the range for $\beta$ where the epidemic is considered localized, and how is this range affected by the structure?

An informal definition is to consider an epidemic localized for $\beta \in [\beta_\mathrm{c}, \beta^*]$, where $\beta^* \equiv \Omega(g_m,p_n;\nu)^{-1}$ acts as a \textit{delocalization threshold}.
Indeed, in the delocalized regime, we have that $\beta_\mathrm{c} \approx \Omega(g_m,p_n;\nu)^{-1}$.
This reinforces the interpretation of $\Omega(g_m,p_n;\nu)$ as a structural \textit{coupling} between the groups: for $\beta \, \Omega(g_m,p_n;\nu) > 1$, the disease is able to efficiently spread between groups, and the disease is sustained \textit{collectively}.

It is analogous to the observation made in Ref.~\cite{mata2015} that the epidemic threshold predicted by the heterogeneous mean-field theory seems to predict the delocalization threshold. In fact, if we take $p_n = \delta_{n,2}$ (equivalent to having configuration model networks with degree distribution $g_m$), the threshold $\beta^*$ is equivalent to the one predicted by pair heterogeneous mean-field theory \cite{mata2014heterogeneous}, i.e.,
\begin{align*}
    \beta^* = \frac{\langle m \rangle}{\langle m (m-1) \rangle} \;.
\end{align*}

In Fig.~\ref{fig:local_scaling}, the shaded region highlights the localized portion of the bifurcation diagram. Note that the right-hand side of this region roughly corresponds to the inflexion point of the global prevalence.


%

\end{document}